\documentclass[sii]{ipart}

\RequirePackage[OT1]{fontenc}
\usepackage{amsfonts,amstext,amsmath,amssymb,amsthm}
\usepackage{dsfont}

\usepackage{rotating}

\usepackage{subfig}

\usepackage{paralist}

\usepackage[numbers,square]{natbib}
\RequirePackage[colorlinks]{hyperref}

\pubyear{2015}
\volume{0}
\issue{0}
\firstpage{1}
\lastpage{14}
\arxiv{1509.01570}

\startlocaldefs

\renewcommand{\Pr}{\mathbb{P}} 
\DeclareMathOperator{\EV}{\mathbb{E}} 

\DeclareMathOperator{\LLR}{\mathcal{L}}
\DeclareMathOperator{\LR}{\Lambda}

\DeclareMathOperator{\ADD}{ADD}
\DeclareMathOperator{\Var}{Var}

\DeclareMathOperator{\ARL}{ARL}
\DeclareMathOperator{\SADD}{SADD}
\DeclareMathOperator{\STADD}{STADD}

\DeclareMathOperator*{\argmax}{arg\,max}

\DeclareMathOperator{\ONE}{\mathchoice{\rm 1\mskip-4.2mu l}{\rm 1\mskip-4.2mu l}{\rm 1\mskip-4.6mu l}{\rm 1\mskip-5.2mu l}}
\newcommand{\indicator}[1]{{\ONE_{\left\{#1\right\}}}}

\newcommand{\T}{T}

\newcommand{\abs}[1]{\left\vert#1\right\vert}

\endlocaldefs

\graphicspath{{./gfx/}}

\begin{document}

\begin{frontmatter}

\title{Real-time financial surveillance via quickest change-point detection methods}

\runtitle{Financial surveillance via change-point detection methods}


\begin{aug}
\author{\fnms{Andrey} \snm{Pepelyshev}\corref{}%
\ead[label=e1]{andrey@ap7236.spb.edu}}
\address{Faculty of Mathematics\\
St.~Petersburg State University\\
Peterhof, St. Petersburg 198504\\
Russia\\[2mm]
\noindent School of Mathematics\\
Cardiff University\\
Cardiff, CF24 4AG\\
UK\\
\printead{e1}}
\and
\author{\fnms{Aleksey S.} \snm{Polunchenko}
\ead[label=e2]{aleksey@binghamton.edu}%
\ead[label=u2,url]{http://www.math.binghamton.edu/aleksey}}
\address{Department of Mathematical Sciences\\
State University of New York at Binghamton\\
Binghamton, New York 13902--6000\\
USA\\
\printead{e2}\\
\printead{u2}}

\runauthor{A.~Pepelyshev and A.S.~Polunchenko}

\end{aug}

\begin{abstract}
We consider the problem of efficient financial surveillance aimed at ``on-the-go'' detection of structural breaks (anomalies) in ``live''-monitored financial time series. With the problem approached statistically, viz. as that of {\em multi-cyclic} sequential (quickest) change-point detection, we propose a semi-parametric multi-cyclic change-point detection procedure to promptly spot anomalies as they occur in the time series under surveillance. The proposed procedure is a derivative of the likelihood ratio-based Shiryaev--Roberts (SR) procedure; the latter is a quasi-Bayesian surveillance method known to deliver the fastest (in the multi-cyclic sense) speed of detection, whatever be the false alarm frequency. We offer a case study where we first carry out, step by step, a preliminary statistical analysis of a set of {\em real-world} financial data, and then set up and devise\begin{inparaenum}[\itshape(a)]\item the proposed SR-based anomaly-detection procedure and \item the celebrated Cumulative Sum (CUSUM) chart \end{inparaenum} to detect structural breaks in the data. While both procedures performed well, the proposed SR-derivative, conforming to the intuition, seemed slightly better.
\end{abstract}

\begin{keyword}[class=AMS]
\kwd[Primary ]{62L10}
\kwd{62L15}
\kwd[; secondary ]{62P05}
\end{keyword}

\begin{keyword}
\kwd{CUSUM chart}
\kwd{Financial surveillance}
\kwd{Sequential analysis}
\kwd{Shiryaev--Roberts procedure}
\kwd{Quickest change-point detection}
\end{keyword}

\received{\smonth{1} \syear{2015}}


\end{frontmatter}

\section{Introduction}
\label{sec:intro} 

The world's history of economic crises, including the latest and still-ongoing global financial meltdown and recession that started in 2008--2009, provides graphic evidence of the importance of efficient methods for continuous financial surveillance~\cite{Frisen:Book2008,Frisen:SA2009}. By allowing to detect anomalous patters {\em early} and {\em reliably}, such methods form a foundation for {\em active} risk management~\cite{Lai+Xing:Book2015}. This paper examines the possibility of approaching the problem of financial monitoring statistically. Specifically, the principal idea is to exploit the machinery of sequential (quickest) change-point detection. The subject is concerned with the development and evaluation of ``watch dog''-type of procedures for early yet reliable detection of unanticipated changes (structural breaks) that may occur in the statistical profile of a ``live''-monitored time series. For an introduction into the subject, see, e.g.,~\cite{Shiryaev:Book78,Zhigljavsky+Kraskovsky:Book1988,Basseville+Nikiforov:Book93,Poor+Hadjiliadis:Book09,Veeravalli+Banerjee:AP2013}, or~\cite[Part~II]{Tartakovsky+etal:Book2014}, and the references therein.

One of the first comprehensive expositions of nonparametric change-point detection methods for {\em online} financial surveillance was offered by Brodsky and Darkhovsky~\cite{Brodsky+Darkhovsky:Book1993,Brodsky+Darkhovsky:Book2000}. More recently, the machinery of Singular Spectrum Analysis (SSA) has also been utilized in~\cite{Golyandina+etal:Book2001,Moskvina+Zhigljavsky:CommStat2003,Zhigljavsky:IWSM2009,Golyandina+Zhigljavsky:Book2013}. In particular, it was demonstrated via numerous case studies involving intricate real-world data that the SSA-based version of Page's~\cite{Page:B54} celebrated Cumulative Sum (CUSUM) ``inspection scheme'' is able to efficiently detect changes of rather complicated structure (e.g., in the frequency of a periodic component of the time series of interest).

However, nearly all of the research on the subject done to date revolves around only three change-point detection methods: the Shewhart $\bar{X}$-chart~\cite{Shewhart:JASA1925,Shewhart:Book1931}, the CUSUM ``inspection scheme''~\cite{Page:B54}, and the Exponentially Weighted Moving Average (EWMA) chart~\cite{Roberts:T59}. Over the years, the three have {\it de facto} become {\em  the} detection tools in applied sequential analysis, especially in quality control. Part of the reason is the methods' simplicity, and another part is their theoretically established strong optimality properties~\cite{Moustakides:AS86,Ritov:AS90,Pollak+Krieger:SA2013}. By contrast, the focus of this paper is on the Shiryaev--Roberts (SR) procedure~\cite{Shiryaev:SMD61,Shiryaev:TPA63,Roberts:T66,Shiryaev:Book78}. Although the SR procedure is only slightly ``younger'' than the CUSUM and EWMA charts, it has heretofore been largely neglected by practitioners as well as by statisticians. Consequently, examples of applications of the SR procedure to real-world data are extremely rare. However, the SR procedure has been recently discovered~\cite{Pollak+Tartakovsky:ISITA2008,Pollak+Tartakovsky:SS09,Shiryaev+Zryumov:Khabanov2010} to possess strong optimality properties in Shiryaev's~\cite{Shiryaev:SMD61,Shiryaev:TPA63,Shiryaev:Book78} multi-cyclic setting, which is a setting adequate in many real-world applications. Motivated by this, the authors of~\cite{Polunchenko+etal:SA2012,Tartakovsy+etal:IEEE-JSTSP2013} have successfully applied the SR procedure in the area of cybersecurity, namely for online detection of anomalies (caused, e.g., by intrusions) in computer networks. The present paper is intended to provide yet another example of an SR-type anomaly-detection algorithm capable of operating on real-world {\em financial} data. Due to the exact multi-cyclic optimality of the SR procedure, the proposed algorithm is expected to compare favorably to other detection schemes, in particular the multi-cyclic CUSUM procedure.

We would like to remark that, to the best of our knowledge, the only other attempt to apply the SR procedure to {\em real-world financial data} would be that made previously by Ergashev~\cite{Ergashev:EconWPA2004}. Specifically, Ergashev~\cite{Ergashev:EconWPA2004} was concerned with the problem of early detection of the ``turning points'' in the US business cycles. These cycles, also known as the US economic cycles, are alternating periods of recession and recovery, manifested in fluctuations of the US economic activity around its long-term potential level. Hence, the ``turning points'' effectively signify the onset of either recession (contraction) or recovery (expansion) of the US economy. To detect these ``turning points'', Ergashev~\cite{Ergashev:EconWPA2004} applied the SR procedure and the CUSUM and EWMA charts to the series of Composite Leading Indicators (CLIs); the CLIs are updated monthly by the Organisation for Economic Co-operation and Development (OECD; see on the Web at~\url{http://www.oecd.org}) to provide early signals of ``turning points'' in the US business cycles. Through experiments involving the actual CLIs series, Ergashev~\cite{Ergashev:EconWPA2004} demonstrated the SR procedure to be better (i.e., quicker) at detecting the US business cycles' ``turning points'' than the CUSUM and EWMA charts with the same level of the ``false positive'' risk. In this work we too provide experimental evidence that the SR procedure might be superior to the CUSUM chart when it comes to detecting structural breaks in time series of {\em real-world} stock prices.

The rest of the paper is organized as follows. We start in Section~\ref{sec:preliminaries} with a brief introduction to the area of quickest change-point detection and provide a short overview of the state-of-the-art in the field. Next, in Section~\ref{sec:application} we offer an SR-based anomaly-detection algorithm suitable to operate on real-world data. Section~\ref{sec:case-study} is devoted to a case study where we devise the proposed algorithm to perform anomaly-detection in a {\em real-world} financial time series. The conclusions follow in Section~\ref{sec:conclusion} which sums up the entire paper.

\section{Preliminary background on quickest change-point detection}
\label{sec:preliminaries}

The aim of this section is two-fold:\begin{inparaenum}[\itshape(a)]\item to provide a short but formal introduction to the problem of quickest change-point detection and \item to give a brief account of the state-of-the-art in the field\end{inparaenum}. This is necessary as background for the later sections. For lack of space, we shall only consider the basic {\em iid} version of the quickest change-point detection problem. For a thorough treatment the general non-iid case, see, e.g.,~\cite{Shiryaev:Book78,Tartakovsky+Moustakides:SA10,Polunchenko+Tartakovsky:MCAP2012} or~\cite[Part~II]{Tartakovsky+etal:Book2014}.

Suppose one is able to {\em sequentially} observe a time series, $\{X_n\}_{n\ge1}$, where $X_i$'s are independent. Suppose further that the statistical structure of the series is such that $X_1,\ldots,X_\nu$ are each distributed according to a known probability density function (pdf) $f(x)$, while $X_{\nu+1},X_{\nu+2},\ldots$ each have a pdf $g(x)\not\equiv f(x)$, also known. The basic {\em iid} quickest change-point detection problem is to detect, as one gathers more and more data, that the baseline pdf of the data is no longer $f(x)$, and do so in an optimal manner. The challenge is that the time index $\nu$, which is referred to as the change-point, is not known in advance and may take place at any time $0\le\nu\le\infty$; here and onward, the notation $\nu=0$ ($\nu=\infty$) is to be understood as the case when the change is in effect from the get-go (or never, respectively). The minimax version of the problem assumes that $\nu$ is unknown (but not random). This is different from the Bayesian version of the problem which regards $\nu$ as random~\cite{Shiryaev:SMD61,Shiryaev:TPA63,Shiryaev:Book78}. In this work we shall focus only on the minimax case.

Statistically, the problem is to sequentially test the hypotheses $\mathcal{H}_k\colon\nu=k$, $0\le k<\infty$ (i.e., that the pdf of the observations changes at epoch $k$) against the alternative hypothesis $\mathcal{H}_\infty\colon\nu=\infty$ (i.e., that the pdf never changes); note that $\mathcal{H}_i\cap\mathcal{H}_j=\varnothing$, $i\neq j$, and that $\cup_{j\ge0}\mathcal{H}_j=\Omega$.

The first step to test $\mathcal{H}_k$ against $\mathcal{H}_\infty$ is to construct the corresponding likelihood ratio (LR). To that end, assuming $X_1,X_2,\ldots,X_n$ have been sampled, the LR is of the form
\begin{equation*}
\LR_{k:n}
\triangleq
\prod_{j=k+1}^n\LR_j,\;\text{where}\;\LR_j\triangleq\frac{g(X_j)}{f(X_j)}
\end{equation*}
for $k<n$ and $\LR_{k:n}\equiv 1$ for $k\ge n$; the latter condition merely means that the change has not yet happened. The sequence $\{\LR_{k:n}\}_{1\le k\le n}$ has to be updated ``on-the-go'' incorporating new data points as they become available.

\begin{figure}[!htb]
    \centering
    \includegraphics[width=0.5\textwidth]{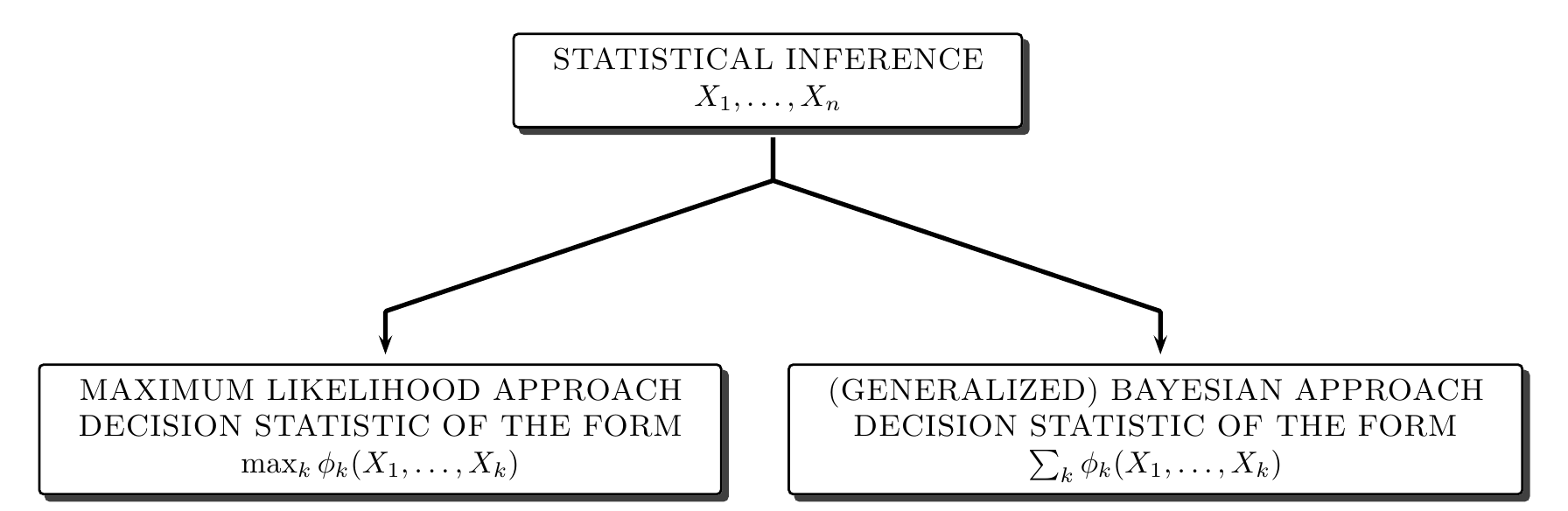}
    \caption{Two different approaches to statistical inference: maximum likelihood and (generalized) Bayesian.}
    \label{fig:stat-inference-appr}
\end{figure}

Once constructed, the LR is turned into a {\em detection statistic} to be subsequently used for actual decision-making. Basing the detection statistic on the LR ensures that the former is sensitive to whether the sample drawn so far is statistically homogeneous or not. There are generally two fundamentally different ways to utilize the LR to design a ``good'' detection statistic: either exploit the maximum likelihood principle or take the (generalized) Bayesian approach. This is shown schematically in Figure~\ref{fig:stat-inference-appr}.

The idea of the maximum likelihood approach is to sequentially maximize $\{\LR_{k:n}\}_{1\le k\le n}$ with respect to the change-point $\nu=k$, where $k=1,2,\ldots,n$. Specifically, the corresponding detection statistic is
\begin{equation}\label{eq:CUSUM_V}
V_n
\triangleq
\max_{1\le k\le n}\LR_{k:n},\; n\ge1,
\end{equation}
which is the famous CUSUM statistic~\cite{Page:B54}. We note that the maximization with respect to $k$ in the right-hand side of~\eqref{eq:CUSUM_V} is possible because the change-point $\nu=k$ is assumed {\em unknown} (nonrandom).

By contrast, the Bayesian approach treats the change-point as a random number, possessing a certain prior distribution~\cite{Girschick+Rubin:AMS52,Shiryaev:SMD61,Shiryaev:TPA63,Tartakovsky+Moustakides:SA10,Polunchenko+Tartakovsky:MCAP2012}. However, since we agreed to assume that $\nu$ is unknown (nonrandom), the corresponding quasi-Bayesian (or generalized Bayesian) detection statistic can be defined as
\begin{equation}\label{eq:Bayesian}
R_n
\triangleq
\sum_{k=1}^n\LR_{k:n},\; n\ge1,
\end{equation}
i.e., $R_n$ is effectively the average of $\{\LR_{k:n}\}_{1\le k\le n}$ taken with respect to the change-point $\nu=k$, $1\le k\le n$ assuming that it follows an (improper) uniform prior distribution; see, e.g.,~\cite{Girschick+Rubin:AMS52,Shiryaev:SMD61,Shiryaev:TPA63,Tartakovsky+Moustakides:SA10,Polunchenko+Tartakovsky:MCAP2012}.

Statistics~\eqref{eq:CUSUM_V} and~\eqref{eq:Bayesian} are the two main choices in all of quickest change-point detection. Both lead to efficient sequential detection procedures. Specifically, a sequential detection procedure is identified with a stopping time, $\T$, which is a functional of the observed data, $\{X_n\}_{n\ge1}$. The meaning of $\T$ is that after observing $X_1,\ldots,X_T$ it is declared that apparently the change is in effect. This need not be the case, and if it is not the case, then $\T\le\nu$ and the detection procedure $\T$ is said to have sounded a false alarm. A ``good'' (i.e., optimal or nearly optimal) detection procedure is one that minimizes (or nearly minimizes) the desired detection delay penalty-function, subject to a constraint on the false alarm risk. For an overview of the major optimality criteria, see, e.g.,~\cite{Tartakovsky+Moustakides:SA10,Polunchenko+Tartakovsky:MCAP2012,Polunchenko+etal:JSM2013,Veeravalli+Banerjee:AP2013}, or~\cite[Part~II]{Tartakovsky+etal:Book2014}.

Let $\Pr_{k}(\cdot)$, $0\le k\le\infty$, denote the probability measure assuming that $\nu=k$, $0\le k\le\infty$ (so that $\Pr_{\infty}(\cdot)$ corresponds to the case when $\nu=\infty$). Let $\EV_{k}[\cdot]$, $0\le k\le\infty$, be the corresponding expectation.

Page~\cite{Page:B54} and then also Lorden~\cite{Lorden:AMS71} proposed to measure the ``false alarm'' risk through the Average Run Length (ARL) to false alarm $\ARL(\T)\triangleq\EV_{\infty}[\T]$. This metric captures the average number of observations that the procedure samples before it triggers a false alarm. The higher (lower) the level of the ARL to false alarm, the lower (higher) the actual level of the ``false alarm'' risk.

A practical approach to quantify the detection speed is to use the ``worst-case'' (Supremum) Average Delay to Detection (ADD),
conditional on a false alarm not having been previously occurred, i.e.,
\begin{equation*}
\SADD(\T)
\triangleq
\max_{0\le k<\infty}\ADD_k(\T),
\end{equation*}
where $\ADD_k(\T)\triangleq\EV_k[\T-k|\T>k]$, $0\le k<\infty$. This metric was introduced by Pollak~\cite{Pollak:AS85}.

Let
\begin{equation*}
\Delta(\gamma)
\triangleq
\Bigl\{\T\colon\ARL(\T)\ge\gamma\Bigr\},\;\gamma>1,
\end{equation*}
i.e., be the class of procedures with the ARL to false alarm of at least $\gamma>1$, an {\it a~priori} chosen level. Then Pollak's~\cite{Pollak:AS85} minimax quickest change-point detection problem is to find $\T_{\mathrm{opt}}\in\Delta(\gamma)$ such that $\SADD(\T_{\mathrm{opt}})=\inf_{\T\in\Delta(\gamma)}\SADD(\T)$ for all $\gamma>1$. This problem is still an open one, and although there has been a continuous effort to solve it, the exact solution has been obtained in only two special cases (see~\cite{Polunchenko+Tartakovsky:AS10,Tartakovsky+Polunchenko:IWAP10}) and, in general, only asymptotic (as $\gamma\to\infty$) solutions have been obtained so far~\cite{Pollak:AS85,Tartakovsky+etal:TPA2012}.

As was mentioned earlier, Page's~\cite{Page:B54} CUSUM chart has been one of the main tools for change-point detection. Part of the reason is the fact that the CUSUM chart is strictly minimax with respect to Lorden's~\cite{Lorden:AMS71} criterion for every $\gamma>1$; see~\cite{Moustakides:AS86,Ritov:AS90}. The CUSUM chart is based on the maximum likelihood principle: it iteratively maximizes $\LLR_n\triangleq\log\LR_n$, i.e., the log-likelihood ratio (LLR), with respect to the change-point $\nu$, and stops as soon as the running maximum exceeds a certain threshold. More specifically, the CUSUM chart is based on the statistic $W_n\triangleq\max\{0, \log V_n\}$, where $V_n$ is as in~\eqref{eq:CUSUM_V}. Note that $W_n$ satisfies the recurrence
\begin{equation}\label{eq:Wn-CS-def}
W_n
\triangleq
\max\{0,W_{n-1}+\LLR_n\},\;n\ge1,\; W_0=0.
\end{equation}

The corresponding stopping rule is
\begin{equation}\label{eq:T-CS-def}
\mathcal{C}_h
\triangleq
\min\{n\ge1\colon W_n\ge h\},
\end{equation}
where $h>0$ is a detection threshold preset so as to achieve the desired level $\gamma>1$ of the ARL to false alarm,
and thus guarantee $\mathcal{C}_h\in\Delta(\gamma)$. Since $\ARL(\mathcal{C}_h)\ge e^h$ for any $h>0$ (see~\cite{Lorden:AMS71} for a proof), setting $h=h_\gamma\ge\log\gamma$ is sufficient to ensure $\mathcal{C}_h\in\Delta(\gamma)$. A more accurate approximation (mentioned, e.g., in~\cite{Polunchenko+etal:SA2012}) for $\ARL(\mathcal{C}_h)$ is as follows:
\begin{equation}\label{eq:ARLcusum}
\ARL(\mathcal{C}_{h})
\approx
\frac{e^h}{I_g\zeta^2}-\frac{h}{I_f}-\frac{1}{I_g\zeta},
\end{equation}
where $I_f\triangleq -\EV_{\infty}[\LLR_1]$ and $I_g\triangleq\EV_0[\LLR_1]$ denote the Kullback--Leibler information numbers (here and throughout the rest of this section it is to be assumed that $0<I_f<\infty$ and $0<I_g<\infty$). The indices $I_f$ and $I_g$ that appear in the right-hand side of~\eqref{eq:ARLcusum} are quantitative measures of the ``contrastness'' of the change, and play an important role in change-point detection.

To define $\zeta$, let $\{Z_n\}_{n\ge0}$ be the random walk $Z_n\triangleq\sum_{j=1}^n \LLR_j$, $n\ge1$, with $Z_0=0$. For $a\ge 0$, introduce the one-sided stopping time $\tau_a\triangleq\inf\{n\ge1\colon Z_n\ge a\}$ and let $\kappa_a\triangleq Z_{\tau_a}-a$ denote the overshoot (i.e., the excess of $Z_n$ over the level $a$ at stopping). Then $\zeta\triangleq\lim_{a\to\infty}\EV_0[e^{-\kappa_a}]$, which is the limiting exponential overshoot. This model-dependent constant falls within the scope of nonlinear renewal theory, and it can be shown that
\begin{equation}\label{eq:zeta}
\zeta
=
\frac{1}{I_g}\exp\left\{-\sum_{k=1}^\infty\frac{1}{k}\bigl[\Pr_{\infty}(Z_k>0)+\Pr_0(Z_k\le 0)\bigr]\right\};
\end{equation}
cf.,~e.g.,~\cite[Chapters~2~\&~3]{Woodroofe:Book82} and~\cite[Chapter~VIII]{Siegmund:Book85}.

Define also $\varkappa\triangleq\lim_{a\to\infty}\EV_0[\kappa_a]$, which is the limiting overshoot. By methods of nonlinear renewal theory it can also be shown that
\begin{equation}\label{eq:kappa}
\varkappa
=
\frac{\EV_0[Z_1^2]}{2\EV_0[Z_1]}+\sum_{k=1}^\infty\frac{1}{k}\EV_0[\min\{0,Z_k\}];
\end{equation}
cf.,~e.g.,~\cite[Chapters~2~\&~3]{Woodroofe:Book82} and~\cite[Chapter~VIII]{Siegmund:Book85}. In practice, $\zeta$ and $\varkappa$ are usually computed numerically using~\eqref{eq:zeta} and~\eqref{eq:kappa}, respectively.

It can be shown (see, e.g.,~\cite{Siegmund:Book85}) that for the basic iid change-point problem $\SADD(\mathcal{C}_{h})\equiv \EV_0[\mathcal{C}_{h}]$. Let $h=h_\gamma$, where $h_\gamma$ is the solution of the equation $\ARL(\mathcal{C}_{h_\gamma})=\gamma$. Then
\begin{equation}\label{eq:SADDCS}
\SADD(\mathcal{C}_{h_\gamma})
=
\frac{1}{I_g}(h_\gamma+\varkappa+\beta_0)+o(1)\;\text{as}\;\gamma\to\infty,
\end{equation}
where $\beta_0\triangleq\EV_0[\min_{n\ge0} Z_n]$. This property of the CUSUM chart is known as second order asymptotic $\SADD(\T)$-optimality. Expansion~\eqref{eq:SADDCS} was first obtained in~\cite{Dragalin:PSIM94} for the single-parameter exponential family. However, it holds in a more general case as well, as long as certain mild conditions imposed on $\LLR_1$ are satisfied. See~\cite{Tartakovsky:IEEE-CDC05}, where it is also shown that
\begin{equation} \label{eq:ADDinftyCS}
\lim_{k\to\infty}\ADD_k(\mathcal{C}_{h_\gamma})
=
\frac{1}{I_g}(h_\gamma+\varkappa-\beta_{\infty})+o(1)\;\text{as}\;\gamma\to\infty,
\end{equation}
where $\beta_{\infty}\triangleq\lim_{n\to\infty}\EV_{\infty}[Z_n-\min_{0\le k\le n}Z_k]$. In practice, constants $\beta_0$ and $\beta_{\infty}$ are also usually computed numerically (e.g., by Monte Carlo simulations). We also note that the two asymptotics~\eqref{eq:SADDCS} and~\eqref{eq:ADDinftyCS} are inversely proportional to the Kullback-Leibler information number $I_g$. This number is sensitive to how faint or contrast the change is. Specifically, $I_g$ is small for faint changes, and is large otherwise. Therefore, according to~\eqref{eq:SADDCS} and~\eqref{eq:ADDinftyCS}, the average delay to detection turns out to be large for faint changes and small otherwise, which makes perfect sense.

\begin{figure*}[!t]
    \centering
    \subfloat[An example of the behavior of a process of interest with a change in mean at time $\nu$.]{
        \includegraphics[width=0.9\textwidth]{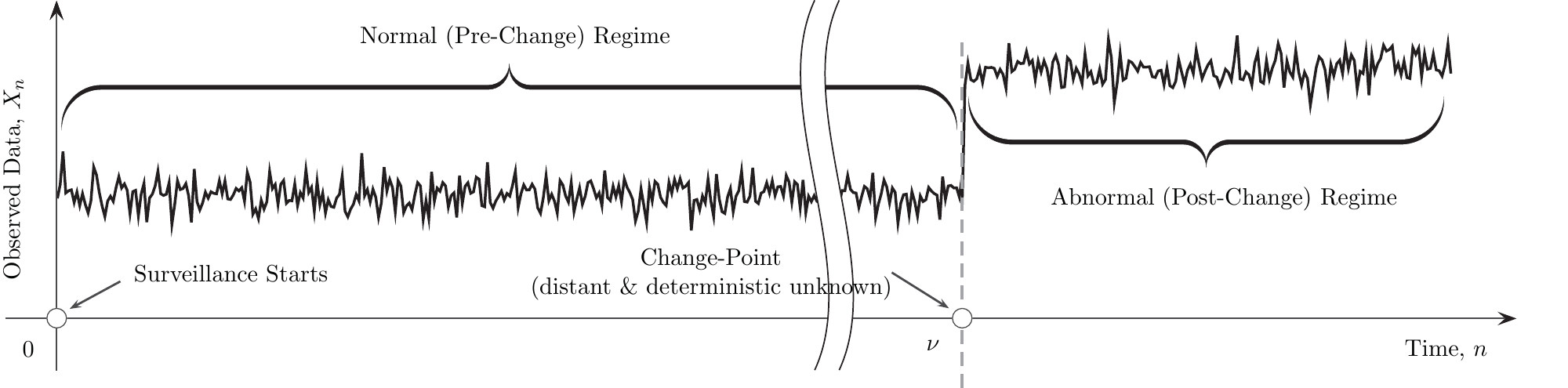}
    } 
    \;\;
    \subfloat[Typical behavior of the detection statistic in the multi-cyclic mode.]{
        \includegraphics[width=0.9\textwidth]{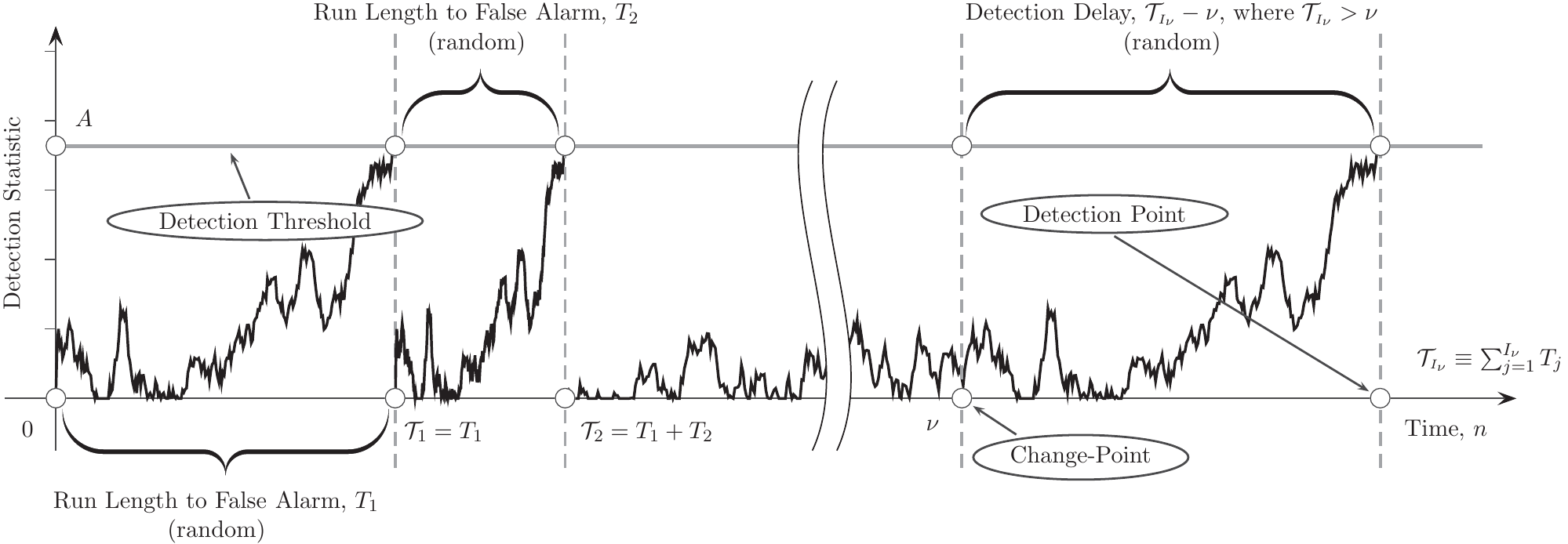}
    } 
    \caption{Multi-cyclic change-point detection in a stationary regime.}
    \label{fig:multi-cyclic-idea}
\end{figure*}

Consider now a context in which it is of utmost importance to detect the change as quickly as possible, even at the expense of raising many false alarms (using a repeated application of the same stopping rule) before the change occurs. Put otherwise, in exchange for the assurance that the change will be detected with maximal speed, we agree to go through a ``storm'' of false alarms along the way (the false alarms are ensued from repeatedly applying the same detection rule, starting from scratch after each false alarm). This scenario is shown in Figure~\ref{fig:multi-cyclic-idea}.

Formally, let $T_1,T_2,\ldots$ be sequential independent repetitions of the stopping time $\T$, and let ${\cal T}_j\triangleq T_1+T_2+\cdots+T_j$, $j\ge1$, be the time of the $j$-th alarm. Define $I_\nu\triangleq\min\{j\ge1\colon {\cal T}_j>\nu\}$. In other words, ${\cal T}_{\scriptscriptstyle I_\nu}$ is the time of detection of a true change that occurs at $\nu$ after $I_\nu-1$ false alarms have been raised. Write
\begin{equation*}
\STADD(\T)
\triangleq
\lim_{\nu\to\infty}\EV_\nu[{\cal T}_{\scriptscriptstyle I_\nu}-\nu]
\end{equation*}
for the limiting value of the average delay to detection referred to as the {\em Stationary Average Delay to Detection} (STADD). The multi-cyclic change-point detection problem is to find $\T_{\mathrm{opt}}\in\Delta(\gamma)$ such that $\STADD(\T_{\mathrm{opt}})=\inf_{\T\in\Delta(\gamma)}\STADD(\T)$ for every $\gamma>1$. Since in this setup $\ARL(\T)$ is effectively the average distance between successive false alarms, the reciprocal $1/\ARL(\T)$ can be interpreted as the frequency of false alarms. The ``intrinsic assumption'' of the multi-cyclic change-point detection problem is that the process under surveillance is not expected to be affected by change ``for a while'', i.e., the change-point, $\nu$, is large. This is a reasonable assumption, e.g., in the area of computer network anomaly detection (see, e.g.,~\cite{Polunchenko+etal:SA2012,Tartakovsy+etal:IEEE-JSTSP2013}) and in financial surveillance.

As has been shown in~\cite{Pollak+Tartakovsky:ISITA2008,Pollak+Tartakovsky:SS09,Shiryaev+Zryumov:Khabanov2010}, the Shiryaev--Roberts (SR) procedure~\cite{Shiryaev:SMD61,Shiryaev:TPA63,Roberts:T66} is {\em exactly} optimal for every $\gamma>1$ with respect to the stationary average detection delay $\STADD(\T)$. Thus, in the multi-cyclic setting the SR procedure is a better alternative to the popular CUSUM chart.

The SR rule calls for stopping at epoch
\begin{equation}\label{eq:T-SR-def}
\mathcal{S}_A
\triangleq
\min\{n\ge1\colon R_n\ge A\},
\end{equation}
where the SR statistic $\{R_n\}_{n\ge0}$ is given by the recursion
\begin{equation}\label{eq:Rn-SR-def}
R_n
=
(1+R_{n-1})\LR_n,\; n\ge1,\; R_0=0;
\end{equation}
cf.~\cite{Shiryaev:SMD61,Shiryaev:TPA63} and~\cite{Roberts:T66}; here $A>0$ is a detection threshold set {\it a~priori} so as to ensure $\mathcal{S}_A\in\Delta(\gamma)$ for a desired $\gamma>1$. It can be easily shown~\cite{Pollak:AS87} that $\ARL(\mathcal{S}_A)\ge A$ for all $A>0$. Hence, setting $A_\gamma=\gamma$ is sufficient to guarantee $\mathcal{S}_{A}\in\Delta(\gamma)$. A more accurate asymptotic approximation is $\ARL(\mathcal{S}_A)\approx A/\zeta$, as $A\to\infty$; see~\cite{Pollak:AS87}.

Let $R_{\infty}$ be a random variable that has the $\Pr_{\infty}$-limiting (stationary) distribution of $R_n$ as $n\to\infty$, i.e., $Q_{\mathrm{ST}}(x)\triangleq\lim_{n\to\infty}\Pr_{\infty}(R_n\le x)=\Pr_{\infty}(R_{\infty}\le x)$. Let $U\triangleq\sum_{k=1}^\infty e^{-Z_k}$ and $\tilde{Q}(x)\triangleq\Pr_0 (U\le x)$.

A straightforward argument shows that, for the SR procedure considered under the basic iid change-point setup, if $A=A_\gamma$ is the solution of the equation $\ARL(\mathcal{S}_{A_\gamma})=\gamma$, then $\SADD(\mathcal{S}_{A_\gamma})
\equiv \EV_0[\mathcal{S}_{A_\gamma}]$, and
\begin{equation}\label{eq:SADDSR}
\SADD(\mathcal{S}_{A_\gamma})
=
\frac{1}{I_g}(\log A_\gamma+\varkappa-C_0)+o(1)\;\text{as}\;\gamma\to\infty,
\end{equation}
where
\begin{equation*}
C_0
\triangleq
\EV[\log(1+{U})]
=
\int_0^\infty\log(1+x)\,d\tilde{Q}(x);
\end{equation*}
cf.~\cite{Tartakovsky+etal:TPA2012}. The asymptotic expansion~\eqref{eq:SADDSR} shows that the SR procedure is also asymptotically second-order $\SADD(\T)$-minimax. In general, constant $C_0$ and distribution $\tilde{Q}(x)$ are amenable to numerical treatment. For cases where both can be computed analytically and in a closed form see~\cite{Tartakovsky+etal:TPA2012} and~\cite{Polunchenko+Tartakovsky:MCAP2012}.

For the multi-cyclic setting we have
\begin{equation*}
\STADD(\mathcal{S}_{A_\gamma})
=
\frac{1}{I_g}(\log A_\gamma+\varkappa -C_\infty)+o(1)\;\text{as}\;\gamma\to\infty,
\end{equation*}
where
\begin{equation*}
\begin{split}
C_{\infty}
&\triangleq
\EV[\log(1+R_{\infty}+{U})]\\
&=
\int_0^\infty\int_0^\infty\log(1+x+y)\,dQ_{\mathrm{ST}}(x)\,d\tilde{Q}(y);
\end{split}
\end{equation*}
cf.~\cite{Tartakovsky+etal:TPA2012}.

We conclude this section with a remark that the exact multi-cyclic optimality property of the SR procedure~\eqref{eq:Rn-SR-def}--\eqref{eq:T-SR-def} depends heavily on the assumption that the pre- and post-change densities $f(x)$ and $g(x)$ are fully known. The consequences of setting up the SR procedure to detect the ``wrong'' change have been recently made clear in~\cite{Du+etal:CommStat2015} where, apparently for the first time in the literature, it was demonstrated experimentally that, if ignored altogether, parametric uncertainty in $g(x)$ may severely affect the STADD delivered by the SR procedure: the relative loss in performance can be on the order of hundreds of percent.

\section{Application to financial surveillance}
\label{sec:application}

Since anomalous events in financial series happen at {\em unknown} points in time, and entail changes in the series' statistical properties, it is intuitively appealing to devise a quickest change-point detection method to detect the onset of such changes as rapidly as possible, while maintaining the false alarm risk at a tolerable level. This section is intended to show how quickest change-point detection can be applied to detect anomalies in ``live'' streams of financial data.

The main difficulty in applying either the CUSUM chart~\eqref{eq:Wn-CS-def}--\eqref{eq:T-CS-def} or the SR procedure~\eqref{eq:Rn-SR-def}--\eqref{eq:T-SR-def} to {\em real-world} data is that the pre- and post-anomaly distributions of the data are poorly understood, if known at all. As a result, any LR-based approach is effectively rendered useless. Hence, a nonparametric approach might be in order. To that end, let us first analyze how the LR exploited by both the CUSUM chart~\eqref{eq:Wn-CS-def}--\eqref{eq:T-CS-def} and the SR procedure~\eqref{eq:Rn-SR-def}--\eqref{eq:T-SR-def} allows the two procedures to sense the presence of a change. To that end, consider the behavior of $\LLR_n\triangleq\log\LR_n$ prior to the change and under the change. Before the change, the LLR has a negative expectation, i.e., $\EV_\infty[\LLR_n]<0$.
This causes the CUSUM statistic to gravitate toward zero in the pre-change regime, and causes the SR statistic to grow slower than it would if the process had already undergone a change. However, as soon as $X_{\nu+1}$---the first ``anomalous'' data point---is recorded, the expectation of the LLR switches its sign to positive, i.e., $\EV_\nu[\LLR_n]>0$, $0\le \nu<n$. As a result, the CUSUM statistic starts to drift away from zero up toward the detection threshold, and the SR statistic's claim rate increases compared to what it would be had there been no change. This difference in the behavior of each one of the two statistics under the pre-change regime and under the post-change regime is the main reason why the CUSUM chart and the SR procedure are able to sense the presence of a change in the observations to begin with.

The above suggests that when it is impossible to construct a LR, the latter can be replaced with a computable score function $S_n\triangleq S_n(X_1,\ldots,X_n)$ such that $\EV_\infty[S_n]<0$ for all $n\ge1$ and $\EV_\nu[S_n]>0$ for all $0\le\nu<n$ with $n\ge1$. This is the key element of the nonparametric approach, and in the context of quickest change-point detection this idea has been previously suggested and explored, e.g., by McDonald~\cite{McDonald:NRL1990}, Lai~\cite{Lai:JRSS95}, Gordon and Pollak~\cite{Gordon+Pollak:AS1994,Gordon+Pollak:AS1995}, and recently also by Pollak~\cite{Pollak:SA2010}. A thorough exposition of the nonparametric approach to change-point detection has been offered by Brodsky and Darkhovsky~\cite{Brodsky+Darkhovsky:Book1993,Brodsky+Darkhovsky:Book2000}.

To be more specific, McDonald~\cite{McDonald:NRL1990} suggested to base surveillance on the series of sequential ranks $U_n\triangleq\sum_{k=1}^n\indicator{X_k<X_n}$ where $\indicator{\cdot}$ denotes the indicator function. The corresponding score function can be taken to be of the form $S_n=U_n-C$, where $C>0$ is a design parameter selected according to the expected type of change and the desired level of the ARL to false alarm. That is, McDonald's~\cite{McDonald:NRL1990} version of the CUSUM chart~\eqref{eq:Wn-CS-def}--\eqref{eq:T-CS-def} signals an alarm according to the stopping time $\mathcal{C}_{h}^{*}=\min\{n\ge1\colon{W}_{n}^{*}\ge h\}$, where $W_{n}^{*}=\max\{0,{W}_{n-1}^{*} + S_n\}$, $n\ge1$, and $h\ge0$ is the detection threshold. If the observations $\{X_n\}_{n\ge1}$ are all iid, then the sequential ranks $U_n$ are approximately uniform, whatever be the observations' common baseline distribution. However, if effective the $\nu$-th data point, $X_{\nu}$, the baseline distribution switches to a stochastically larger distribution, the sequential ranks become larger causing the rank-based CUSUM chart to trigger an alarm. This idea of McDonald~\cite{McDonald:NRL1990} was then extended to the SR procedure by Gordon and Pollak~\cite{Gordon+Pollak:AS1994,Gordon+Pollak:AS1995} and by Pollak~\cite{Pollak:SA2010}.

More generally, for any appropriately designed score function $S_n$, the original SR statistic $\{R_n\}_{n\ge0}$ given by~\eqref{eq:Rn-SR-def} can be replaced with
\begin{equation}\label{eq:Rn-NPSR-def}
\tilde{R}_n
=
(1+\tilde{R}_{n-1})e^{S_n},\; n\ge1, \; \tilde{R}_0=0,
\end{equation}
so that the corresponding SR stopping time is the form
\begin{equation}\label{eq:T-NPSR-def}
\tilde{\mathcal{S}}_{A}
\triangleq
\min\{n\ge1\colon\tilde{R}_n\ge A\},
\end{equation}
where $A>0$ is the detection threshold. Likewise, for the CUSUM chart, the original CUSUM statistic $\{W_n\}_{n\ge0}$ given by~\eqref{eq:Wn-CS-def} can be replaced with
\begin{equation}\label{eq:Wn-NPCS-def}
\tilde{W}_n
=
\max\{0,\tilde{W}_{n-1} + S_n\}, \; n\ge1,\; \tilde{W}_n=0,
\end{equation}
so that the corresponding CUSUM stopping time becomes
\begin{equation}\label{eq:T-NPCS-def}
\tilde{\mathcal{C}}_{h}
\triangleq
\min\{n\ge1\colon\tilde{W}_n\ge h\},
\end{equation}
where $h>0$ is again the detection threshold.

In order for the score-function-based SR procedure~\eqref{eq:Rn-NPSR-def}--\eqref{eq:T-NPSR-def} and CUSUM chart~\eqref{eq:Wn-NPCS-def}--\eqref{eq:T-NPCS-def} to work well, the score function $S_n\triangleq S_n(X_1,\ldots,X_n)$ has to be carefully designed, incorporating the type of change expected. To illustrate this point, suppose we are interested in detecting a change in both the mean and variance of the observations. Let $\mu_\infty\triangleq\EV_\infty[X_n]$ and $\sigma_\infty^2\triangleq\Var_\infty[X_n]$, and $\mu\triangleq\EV_0[X_n]$ and $\sigma^2\triangleq \Var_0[X_n]$ denote the pre- and post-anomaly mean values and variances, respectively. Introduce $\tilde{X}_n\triangleq (X_n-\mu_\infty)/\sigma_\infty$, i.e., the centered and standardized $n$-th data point. In real-world applications the pre-change parameters $\mu_\infty$ and $\sigma_\infty^2$ can usually be estimated in advance (e.g., using training data) and then periodically re-estimated to account for the nonstationary nature of the data. To deal with the uncertainty in $\mu$ and $\sigma^2$, consider the following linear-quadratic score function
\begin{equation}\label{eq:score}
S_n(\tilde{X}_n)
=
C_1 \tilde{X}_n + C_2 \tilde{X}_n^2 - C_3,
\end{equation}
where $C_1$, $C_2$ and $C_3$ are design parameters; cf.~\cite{Tartakovsy+etal:IEEE-JSTSP2013}. Selecting $C_1,C_2$ and $C_3$ to be positive would make this score function sensitive to increases in the mean and variance. In the case when the variance either does not change at all or changes relatively insignificantly compared to the magnitude of the change in the mean, the coefficient $C_2$ may be set equal to zero. This appears to be typical for many cybersecurity applications~\cite{Tartakovsky+etal:SP06,Tartakovsky+etal:SM06+discussion,Tartakovsy+etal:IEEE-JSTSP2013}. In the opposite case when the mean changes only slightly compared to the variance, one may take $C_1=0$.

Note that the score function $S_n$ given by~\eqref{eq:score} with
\begin{equation}\label{eq:design_C}
C_1 = \delta q^2,  \quad  C_2 = \frac{1 - q^2}{2},   \quad C_3 = \frac{\delta^2 q^2}{2} - \log q,
\end{equation}
where $q=\sigma_\infty/\sigma$, $\delta=(\mu-\mu_\infty)/\sigma_\infty$, is optimal if the pre- and post-change distributions are Gaussian with known $\mu$ and $\sigma^2$. This is true because the score function $S_n$ given by~\eqref{eq:score} is then simply nothing but the LLR. If one has reason to believe that the time series of interest can be accurately described by the Gaussian model, then selecting $q=q_0$ and $\delta=\delta_0$ with some design values $q_0$ and $\delta_0$ would lead to decent performance of the procedure for $q < q_0$ and $\delta > \delta_0$ and optimal (i.e., best) performance for $q=q_0$ and $\delta=\delta_0$. However, it is important to emphasize that the proposed score-based ``tweak'' of SR procedure does not require the observations to be Gaussian, whether pre- or post-change.

For examples of score-functions that exploit SSA, see, e.g.,~\cite{Golyandina+etal:Book2001,Moskvina+Zhigljavsky:CommStat2003,Zhigljavsky:IWSM2009,Golyandina+Zhigljavsky:Book2013}.

Another way to deal with parametric uncertainty in the observations' post-change distribution is to employ the generalized likelihood ratio (GLR) approach. However, the obvious problem with this approach is that the recursive evaluation of the running LR---either as in~\eqref{eq:CUSUM_V} or as in~\eqref{eq:Rn-SR-def}---might get computationally too difficult to carry out, because now the LR has to be also maximized with respect to the unknown parameter. As a way around this, Willsky and Jones~\cite{Willsky+Jones:IEEE-AC1976} and then also Lai~\cite{Lai:JRSS95,Lai:IEEE-IT1998} suggested to restrict attention to a certain {\em limited} number of the most recent observations, and, based on that idea, introduced the appropriate ``window-limited'' modification of the CUSUM chart. The main question here, however, is how to choose the size of the window, i.e., the optimal number of the most recent observations to take into account. On the one hand, if that number is too large, the corresponding  ``window-limited'' CUSUM statistic might still be too computationally demanding. On the other hand, basing the decision on too small a number of the latest observations is likely to lead to an increase in the detection delay. To optimize the tradeoff between the computational tractability and the speed of detection, Lai~\cite{Lai:JRSS95,Lai:IEEE-IT1998} showed that the ``best'' strategy is to factor in the latest $M_{\gamma}$ observations with $M_{\gamma}$ being of the order $O(\log\gamma/I_g)$ where $\gamma>1$ is the desired level of the ARL to false alarm and $I_g\triangleq\EV_0[\LLR_1]$ is the Kullback--Leibler information number.


\section{A case study}
\label{sec:case-study}

We now consider a case study where we employ the proposed change-point detection methodology to ``sniff out'' structural breaks in a {\em real-world} financial time series. Specifically, our intent is two-fold: to first provide the steps necessary to configure our change-point detection procedures and then, once the latter are properly set up, to also demonstrate and discuss their performance.

\subsection{Data description}

The time series we would like to study is the daily stock prices (at closing) of Host Hotel \& Resorts, Inc. (see on the Web at~\url{www.hosthotels.com}) for the period from January 3, 2000 through March 30, 2007. Host Hotel \& Resorts, Inc. is the largest American lodging and real estate investment trust (or REIT) headquartered in Bethesda, Maryland, USA. An S\&P 500 and Fortune 500 company, Host Hotel \& Resorts, Inc. is also one of the biggest owners of luxury and upper-upscale hotels. Its hotels are operated under such reputable brand names as Marriott, Ritz--Carlton, Four Seasons, Hyatt and Hilton. Its stock is traded on the New York Stock Exchange (NYSE) under the ticker HST. Our interest in the company is due to its leading position in the industry and the significant size of its assets: as of December 31, 2014, its reported total assets were over \$\,12 billion (with liabilities and debt totaling to about \$\,4.6 billion)~\cite[p.~88]{HST:Form10K-2014}.

Historical data for the HST stock for any period since the stock began trading on the NYSE are freely available on the Internet (e.g., via Yahoo.Finance; see~\url{www.finance.yahoo.com}). We used the Machine Learning Data Set Repository (see on the Web at~\url{www.mldata.org}). The total length of the series is $N=1812$ data points. The choice to focus on the period between January 3, 2000 and March 30, 2007 was because the company's history was very eventful during that time period: the tragic events that took place in New York City on September 11, 2001, and the decade-long global economic unrest that followed caused considerable turbulence in the company's financial well-being. As a result, one would expect the HST stock statistical dynamics within the chosen time frame to experience multiple changes. This makes change-point analysis of the data both interesting and challenging.

\subsection{Preliminary statistical analysis}

\begin{figure}[!htb]
    \includegraphics[width=0.5\textwidth]{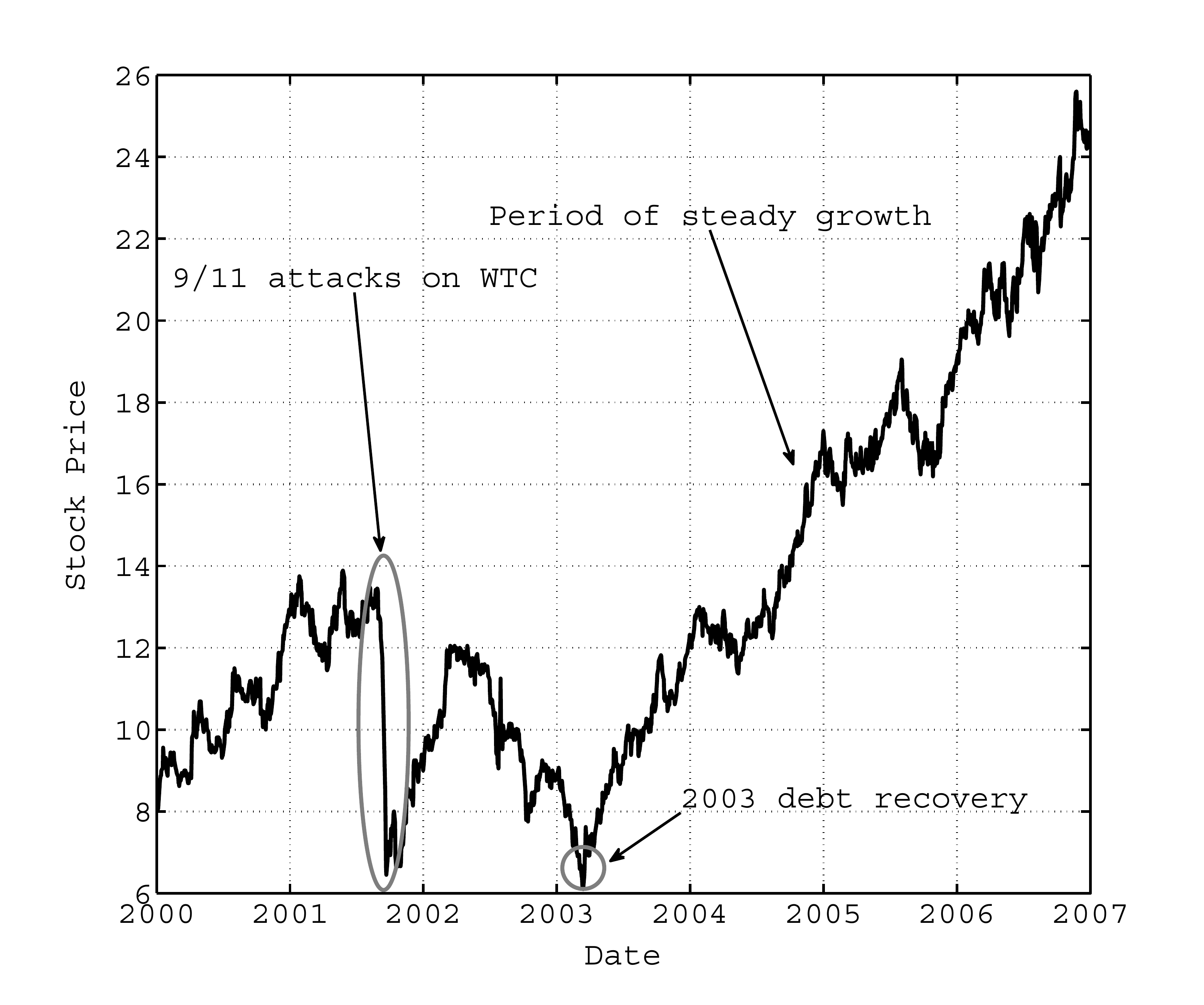}
    \caption{Daily stock prices (at closing) for the Host \& Hotel Resorts, Inc. (NYSE: HST) for the period from January 3, 2003 through March 30, 2007.}
    \label{fig:HST-stock}
\end{figure}

To perform basic statistical analysis of the data, the natural point of departure would be to graph the data against time. This is done in Figure~\ref{fig:HST-stock}. A mere eye examination of the plot suggests several observations.

First note that, as expected, the series appears to be rife with structural breaks of various scale and type. The following three are particularly notable: one occurring toward the end of the third quarter of the year 2001, followed by one more occurring around the end of the first quarter of the year 2002, followed by yet another one occurring in 2003, around the end of the first quarter.

The first of these break-points, viz. the one occurring in 2001, appears to be a crash-type event, as at that point the stock price essentially plummets, from being about \$\,13/share right before the break to being roughly \$\,7/share shortly after the break. The reason for such a huge loss in value is not hard to figure out: it was the result of the 9/11 terrorist attacks on the World Trade Center (WTC) Towers in New York City. Specifically, in addition to destroying the Towers, the attacks also destroyed the New York World Trade Center Marriott hotel owned and operated by Host Hotel \& Resorts, Inc. To boot, the company also sustained considerable damage to its second property located nearby, the New York Marriott Financial Center hotel. However, by the end of 2001, the company received the property and business interruption insurance for the two hotels \cite{HST:Form10K-2001}, and the stock began to claim up.

The second of the above three major break points, namely, the 2002 one, also appears to be a negative event in the Company's history. According to the company's 2002 annual report \cite{HST:AR-2002}, the company's revenue for the year was negatively affected by the overall weakness of the US and global economies, which in particular resulted in business and leisure travel dropping below historic level in 2002.

The third break-point (the one occurring around the first quarter of the year 2003) appears to be a ``turning point'' for the company, because following this break-point, the stock begins to exhibit a consistent upward trend that lasts for years. The specific date of this ``turning point'' is March 14, 2003. According the company's 2003 annual report \cite{HST:AR-2003}, 2003 was indeed a year of recovery for the company: they collected additional insurance on the hotels that were destroyed during the 9/11 attacks in 2001, sold hotels that had been found to be inefficient, and used the proceeds to substantially lessen the corporate debt.

Another observation that can be made from Figure~\ref{fig:HST-stock} is that the HST stock appears to have a seasonal component. This should not come as a surprise, since for the hotel industry seasonal effects are common and, in fact, natural. However, dealing with such effects statistically is somewhat orthogonal to the objective of our study. Nevertheless, we would like to mention that the numerous and extensive case studies offered, e.g., in~\cite{Golyandina+etal:Book2001,Moskvina+Zhigljavsky:CommStat2003,Golyandina+Zhigljavsky:Book2013}, suggest that the SSA methodology can be rather efficient in the analysis of seasonal and cyclic patterns.

\begin{figure}[!htb]
    \centering
    \includegraphics[width=0.5\textwidth]{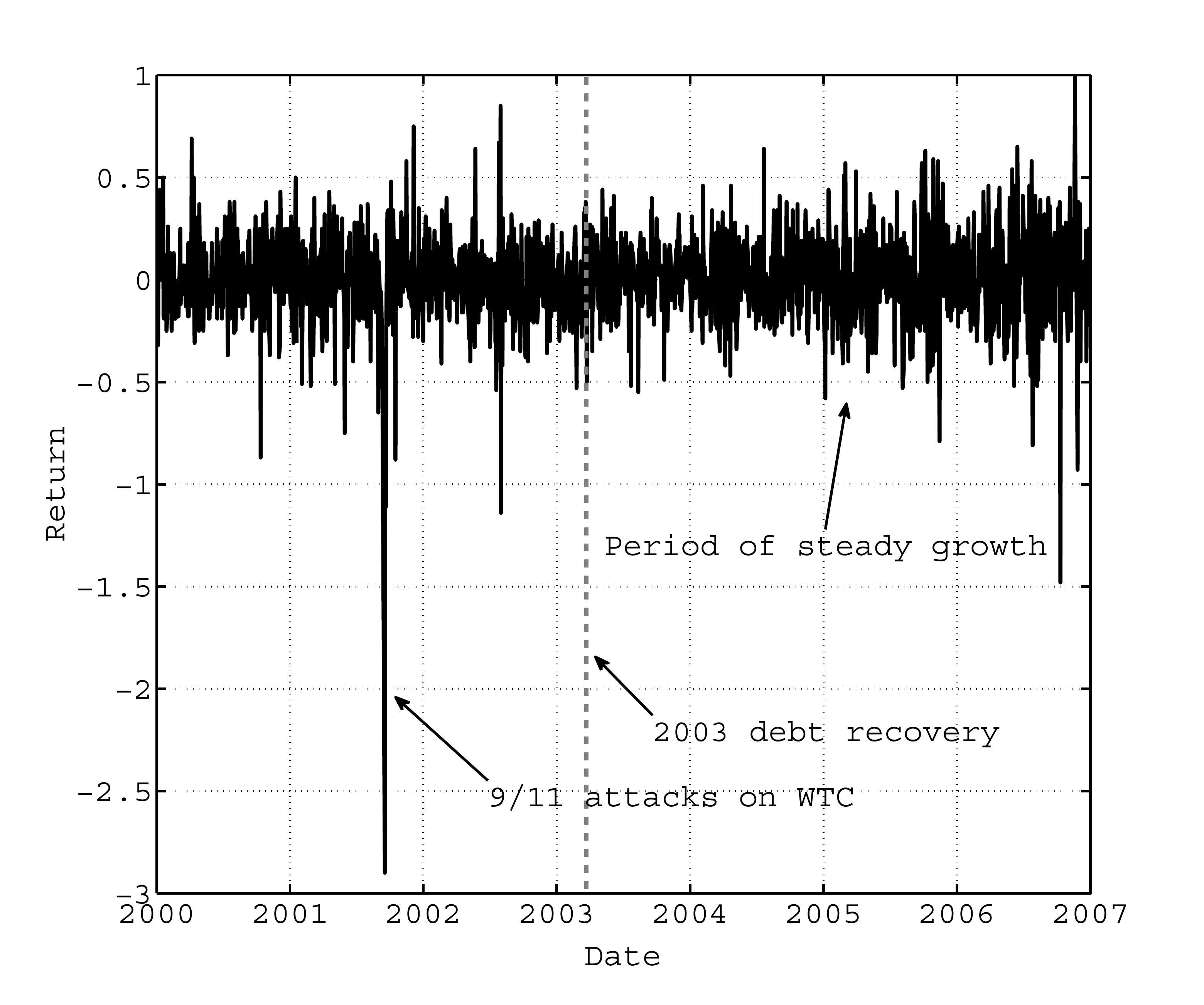}
    \caption{Daily returns on the stock (evaluated at closing) of the Host \& Hotel Resorts, Inc. (NYSE: HST) for the period from January 3, 2003 through March 30, 2007.}
    \label{fig:HST-returns}
\end{figure}

To reinforce the observations made so far, Figure~\ref{fig:HST-returns} shows the behavior of the daily returns $d_i\triangleq X_{i+1}-X_{i}$, $i=1,\ldots,N-1$, on the HST stock. The returns provide a different prospective onto the behavior of the stock itself. As a matter of fact, it is the returns that are usually used as the input data to perform statistical inference on the underlying stock itself. Therefore, we also shall proceed with the returns being the series of interest.

One can clearly see a large down-pointing spike around the third quarter of the year 2001. This spike corresponds to the HST stock loosing approximately half of its value as the result of  the 9/11 terrorist attacks in NYC. While this spike is extremely contrast, there is no apparent change in the daily return distribution corresponding to the 2003 structural break. Nevertheless, as we shall see shortly, the 2003 break-point {\em is} detectable. More importantly, in spite of the steady growth of the stock after the 2003 break-point shown in Figure~\ref{fig:HST-stock}, the behavior of the return series does not confirm any shift in the mean.

\subsection{Offline structural break detection}

We now perform a more thorough statistical analysis of the data. Specifically, we would like to devise a statistical procedure to detect the aforementioned structural breaks. Toward this goal, the first step is to analyze the series {\em retrospectively} so as to not only detect the changes, but also to estimate their locations. One such ``offline'' change-point detection-estimation statistic is the Brodsky--Darkhovsky statistic proposed and studied in~\cite{Brodsky+Darkhovsky:Book1993,Brodsky+Darkhovsky:Book2000}. The Brodsky--Darkhovsky statistic is defined as
\begin{equation}\label{eq:Brodsky-Darkhovsky-stat-def}
Y_N(n)
\triangleq
\sqrt{\frac{n(N-n)}{N^2}}\left[\frac{1}{n}\sum_{i=1}^{n} X_i- \frac{1}{N-n}\sum_{i=n+1}^N X_i\right],
\end{equation}
where $1\le n\le N-1$. As can be seen from the structure of the statistic, it is effectively the difference between two sample means: one computed off the first $n\ge 1$ data points (i.e., $X_1,\ldots,X_n$), and one computed off the remaining $N-n$ data points (i.e., $X_{n+1},\ldots,X_N$) in a chunk of $N\ge n+1$ observations $X_1,\ldots,X_N$. Therefore the statistic~\eqref{eq:Brodsky-Darkhovsky-stat-def} is tailored specifically to detect deviations in the observations' mean. The actual detection procedure consists in comparing  $\abs{Y_N(n)}$ indexed by $n=1,\ldots,N-1$ against a threshold selected according to the desired significance level. More importantly, the statistic can also be used to estimate the actual change-point, $\nu$, i.e., the time moment at which the series' baseline mean (apparently) changes. Specifically, this is accomplished by first identifying the set of values of $n$ for which $\abs{Y_N(n)}$ is maximized, and then using any such $n$ as an estimator of the actual change-point, that is,
\begin{equation}\label{eq:Brodsky-Darkhovsky-est-def}
\hat{\nu}_N
\triangleq
\argmax_{1\le n\le N-1}\abs{Y_N(n)}.
\end{equation}
It has been shown in~\cite{Brodsky+Darkhovsky:Book1993} that such an estimator enjoys strong consistency (as $N\to\infty$) with exponential rate of convergence.

\begin{figure}[!htb]
    \centering
    \includegraphics[width=0.5\textwidth]{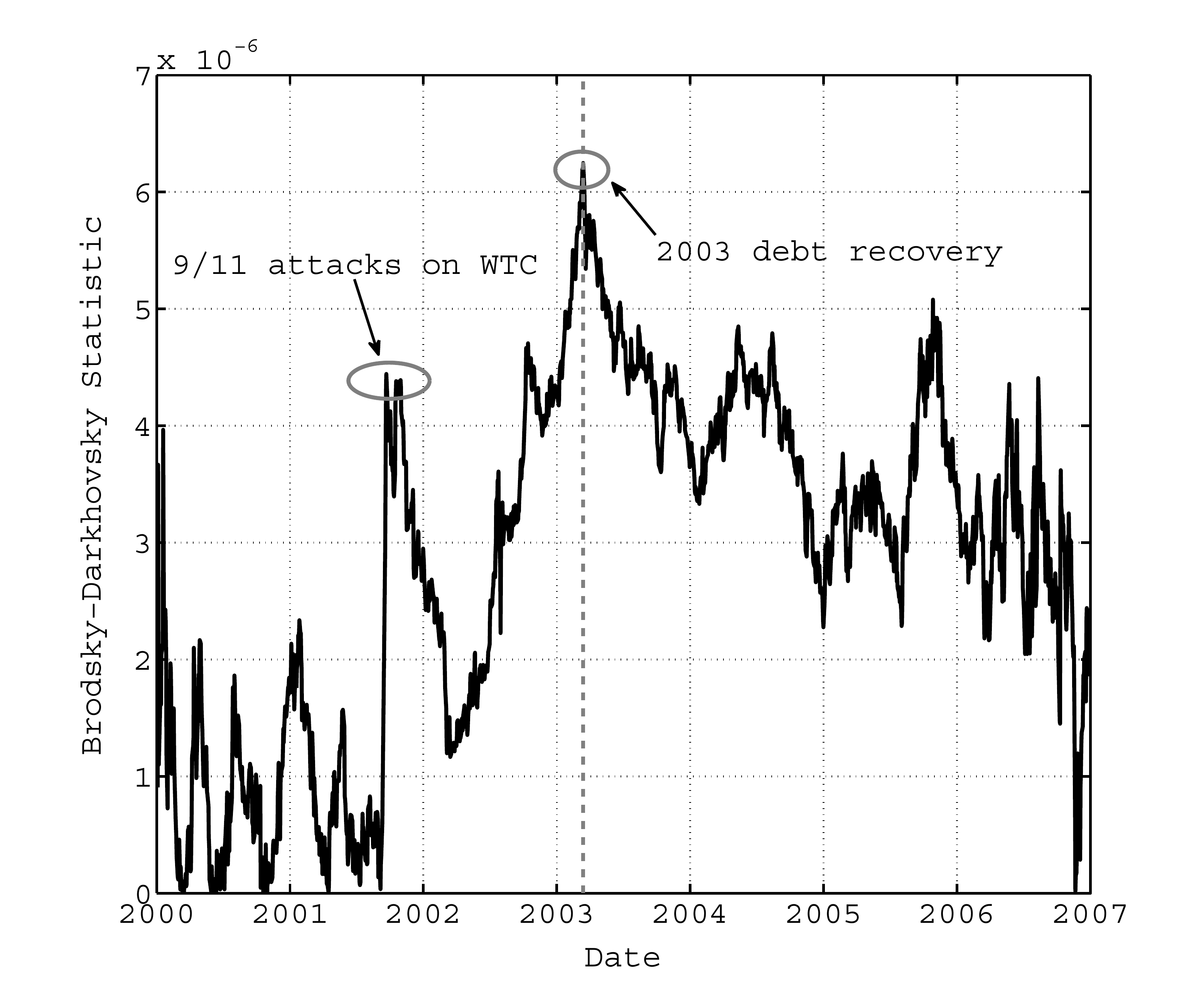}
    \caption{Behavior of the Brodsky--Darkhovsky statistic for the HST stock series.}
    \label{fig:HST-Brodsky-Darkhovsky-stat}
\end{figure}

We have applied the Brodsky--Darkhovsky approach to the returns $\{d_i\}_{1\le i\le N}$, and the obtained behavior of $Y_N(n)$ for $1\le n\le N-1$ is shown in Figure~\ref{fig:HST-Brodsky-Darkhovsky-stat}. It can be seen from the figure that the statistic exhibits a whole series of local yet fairly contrast maxima. The unique and rather strong absolute maximum occurring around the first quarter of 2003 reinforces the observation made earlier that the HST stock undergoes a structural break at that time. The specific location of the absolute maximum corresponds to March 14, 2003, which is the Brodsky--Darkhovsky estimate of the actual change-point. We note that this date is precisely the 2003 break-point.

To continue our analysis of the Brodsky--Darkhovsky statistic shown in Figure~\ref{fig:HST-Brodsky-Darkhovsky-stat}, the spike occurring in the second half of 2001 can be attributed to the 2001 HST stock crash caused by the 9/11 attacks in NYC. The specific value of the Brodsky--Darkhovsky estimate of this change-point is September 18, 2001, which is within the same week of the 9/11 attacks. This estimate can be refined using the following strategy. Once the absolute maximum of the Brodsky--Darkhovsky statistic is identified, the series is partitioned into two nonoverlapping segments: one composed of the observations up to the change-point and one consisting of the observations following the change-point. Then the Brodsky--Darkhovsky detection-estimation method is applied again individually to each of the two data chunks. It is argued in~\cite{Brodsky+Darkhovsky:Book1993,Brodsky+Darkhovsky:Book2000} that this ``divide and conquer'' type of an approach also yields a strongly consistent (as the sample size gets infinitely large) estimator of the change-point.

We now follow this strategy and analyze each piece of data separately. To that end, for the data segment to the left of the 2003 break-point the sample mean and standard deviation are $\hat{\mu}_{\infty}\approx-0.0029$ and $\hat{\sigma}_\infty\approx0.2266$, respectively. The same sample characteristics for data segment to the right of the 2003 break-point turned out to be $\hat{\mu}_{0}\approx 0.0199$ and $\hat{\sigma}_0\approx0.2306$. Therefore, the 2003 break-point changes not only the mean but also the variance. However, the change in the mean is far more contrast than the change in the variance. This could be part of the reason for the excellent performance of the Brodsky--Darkhovsky statistic~\eqref{eq:Brodsky-Darkhovsky-stat-def}.

\begin{figure}[!htb]
    \centering
    \subfloat[Before the 2003 event.]{\label{fig:HST-pre-change-pdf}
        \includegraphics[width=0.5\textwidth]{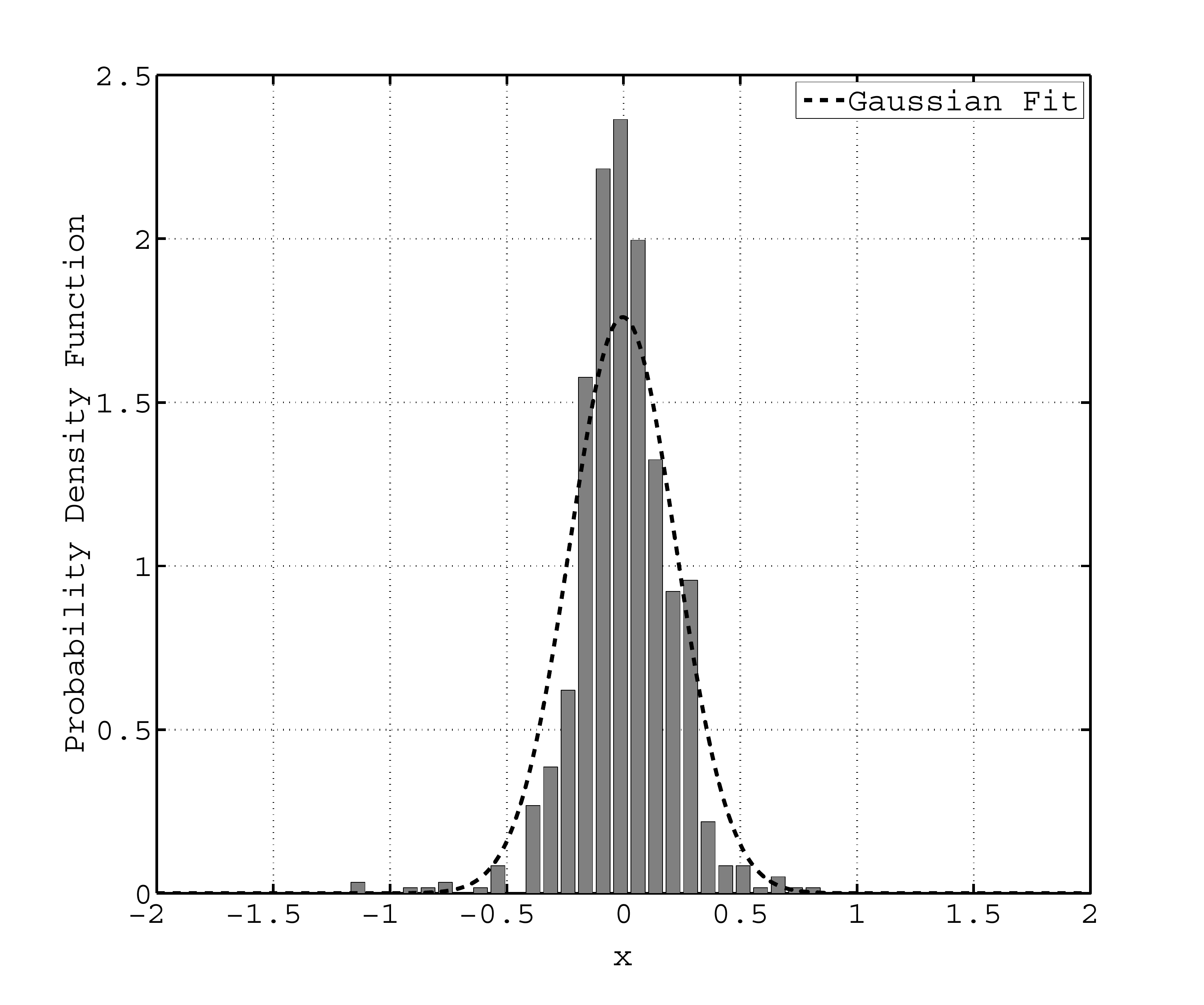}
    }
    \;\;
    \subfloat[After the 2003 event.]{\label{fig:HST-post-change-pdf}
        \includegraphics[width=0.5\textwidth]{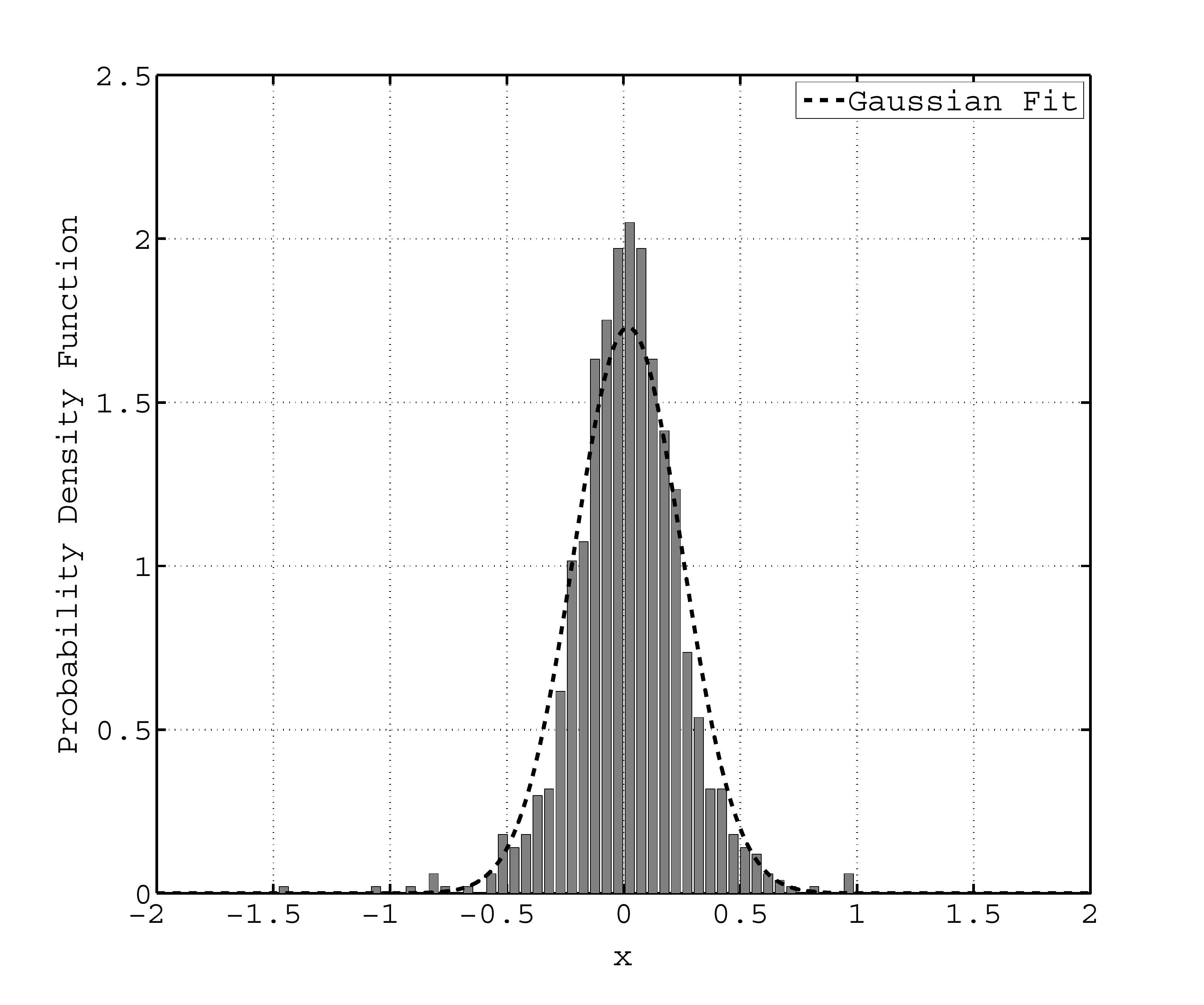}
    }
    \caption{Empirical probability densities for the HST stock returns with Gaussian fits.}
    \label{fig:HST-pdf}
\end{figure}

Figures~\ref{fig:HST-pdf} show the empirical probability densities (histograms) for the returns before (see Figure~\ref{fig:HST-pre-change-pdf}) and after (see Figure~\ref{fig:HST-post-change-pdf}) the 2003 event. Each of the two figures is also accompanied with a Gaussian fit with the mean and variance set to the respective estimated values. Since the two histograms are close to the Gaussian fits, there is only one conclusion to draw: the returns do behave as if they were generated by a Gaussian process.

\begin{figure}[!htb]
    \centering
    \subfloat[Before the 2003 event.]{\label{fig:HST-pre-change-qq}
        \includegraphics[width=0.5\textwidth]{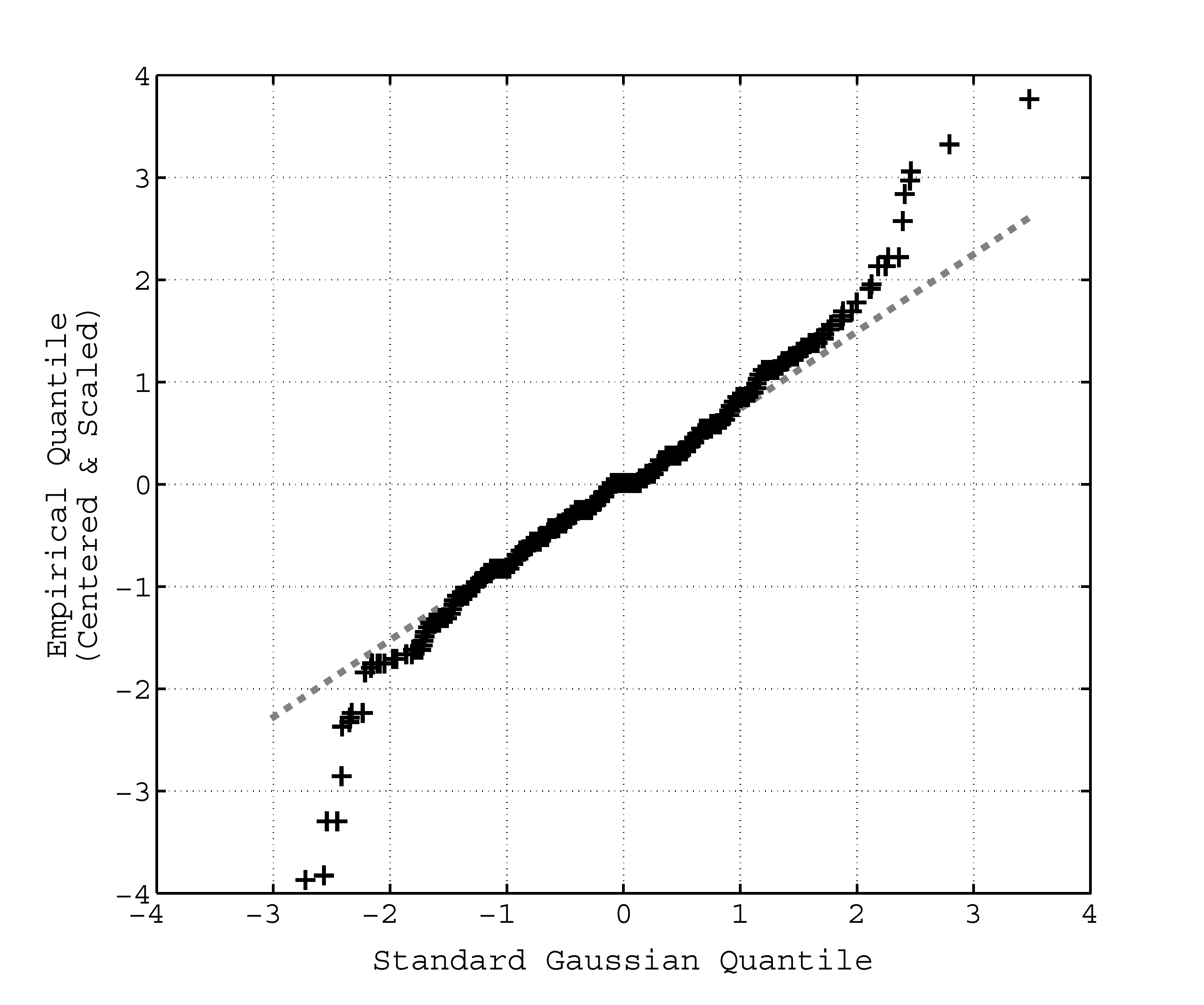}
    }
    \;\;
    \subfloat[After the 2003 event.]{\label{fig:HST-post-change-qq}
        \includegraphics[width=0.5\textwidth]{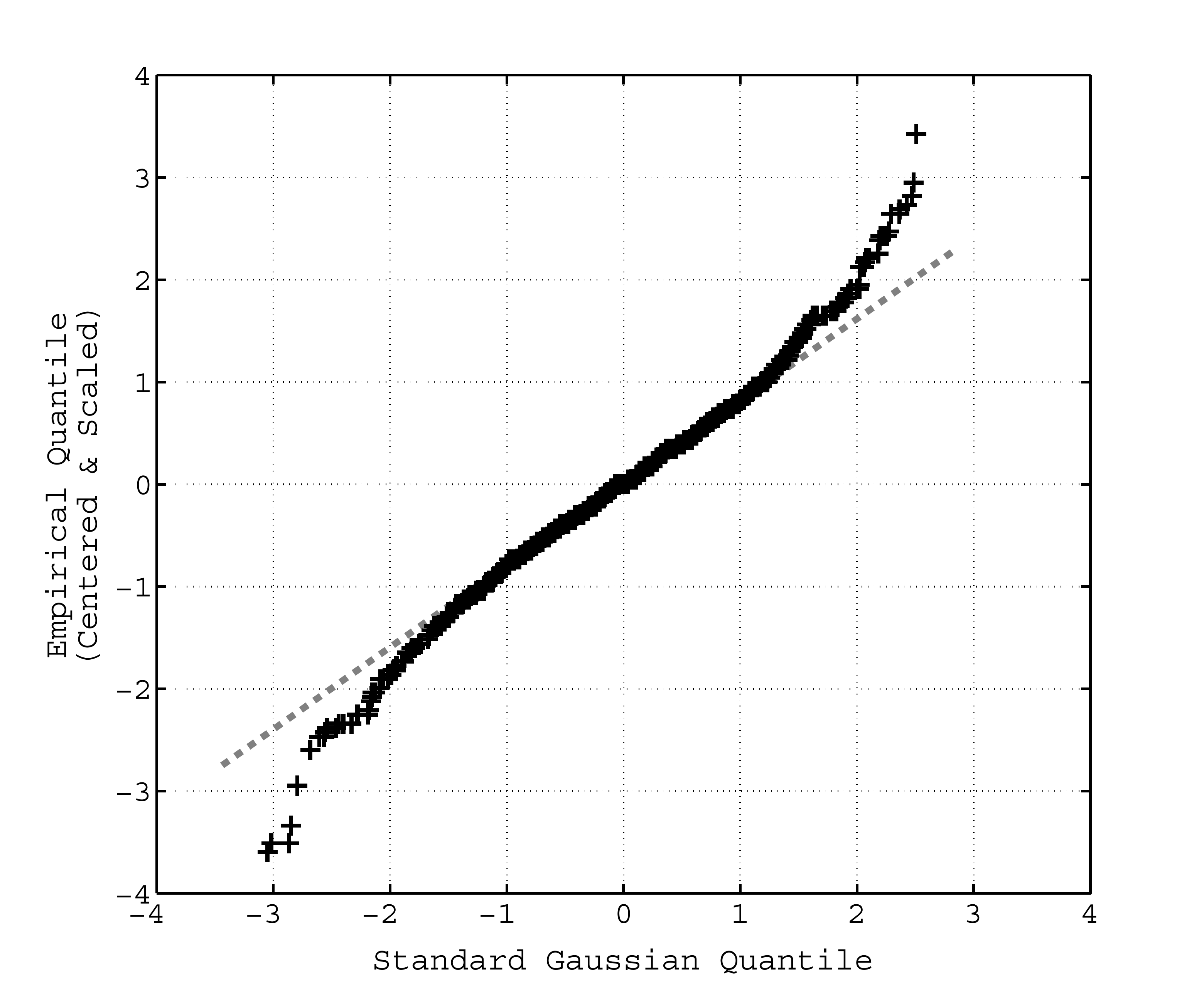}
    }
    \caption{Q-Q plots for the HST stock return distribution vs. the standard Gaussian distribution.}
    \label{fig:HST-QQ}
\end{figure}

The same conclusion can be drawn from an eye inspection of the corresponding Q-Q plots (quantile-quantile) shown in Figure~\ref{fig:HST-QQ}. Specifically, the Q-Q plot for the distribution of the daily returns before the onset of the drift is shown in Figure~\ref{fig:HST-pre-change-qq} and the Q-Q plot for the distribution with the drift in effect is shown in Figure~\ref{fig:HST-post-change-qq}. Since both plots use centered and scaled data, the fitted Gaussian distribution is the standard normal distribution. The fact that both plots are effectively a straight line is evidence of the ``Gaussianness'' of the return distribution before and after the drift.

\begin{figure}[!htb]
    \centering
    \includegraphics[width=0.5\textwidth]{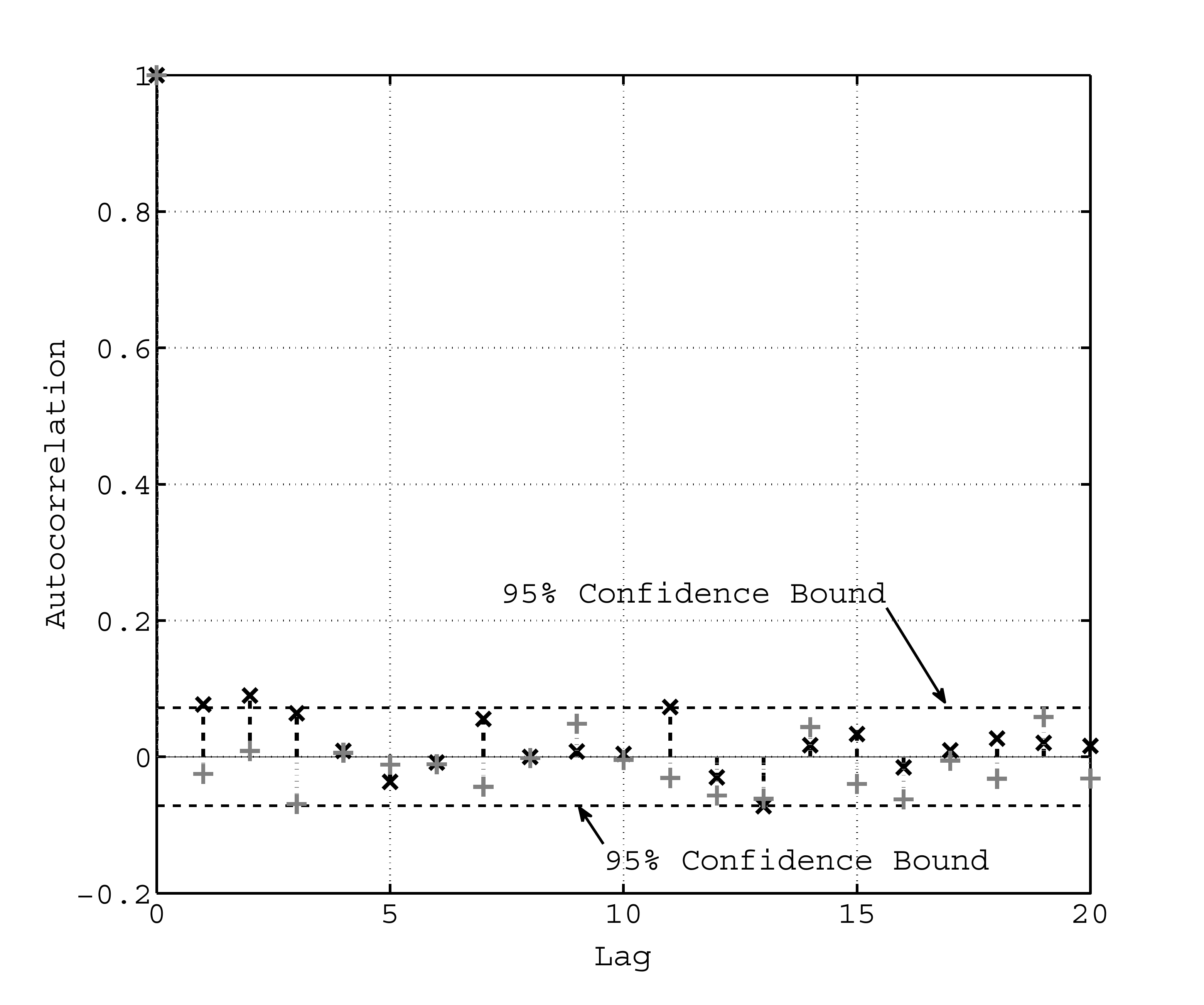}
    \caption{Autocorrelation function for the HST stock returns.}
    \label{fig:HST-autocorr}
\end{figure}

Another important question to be examined about the time series at hand concerns the series' correlation structure. To that end, Figure~\ref{fig:HST-autocorr} shows the correlation plot for the HST stock daily return series. Specifically, the plot distinguishes whether the data are before the 2003 event or after the 2003, and shows the autocorrelation function for the former piece in black and for the latter one in gray. It is clear from the plot that the data are essentially random throughout the entire set, as they exhibit no strong structure or correlation.

\begin{figure*}[!t]
    \centering
    \subfloat[Lag-1 plot.]{\label{fig:HST-lagplot1}
        \includegraphics[width=0.31\textwidth]{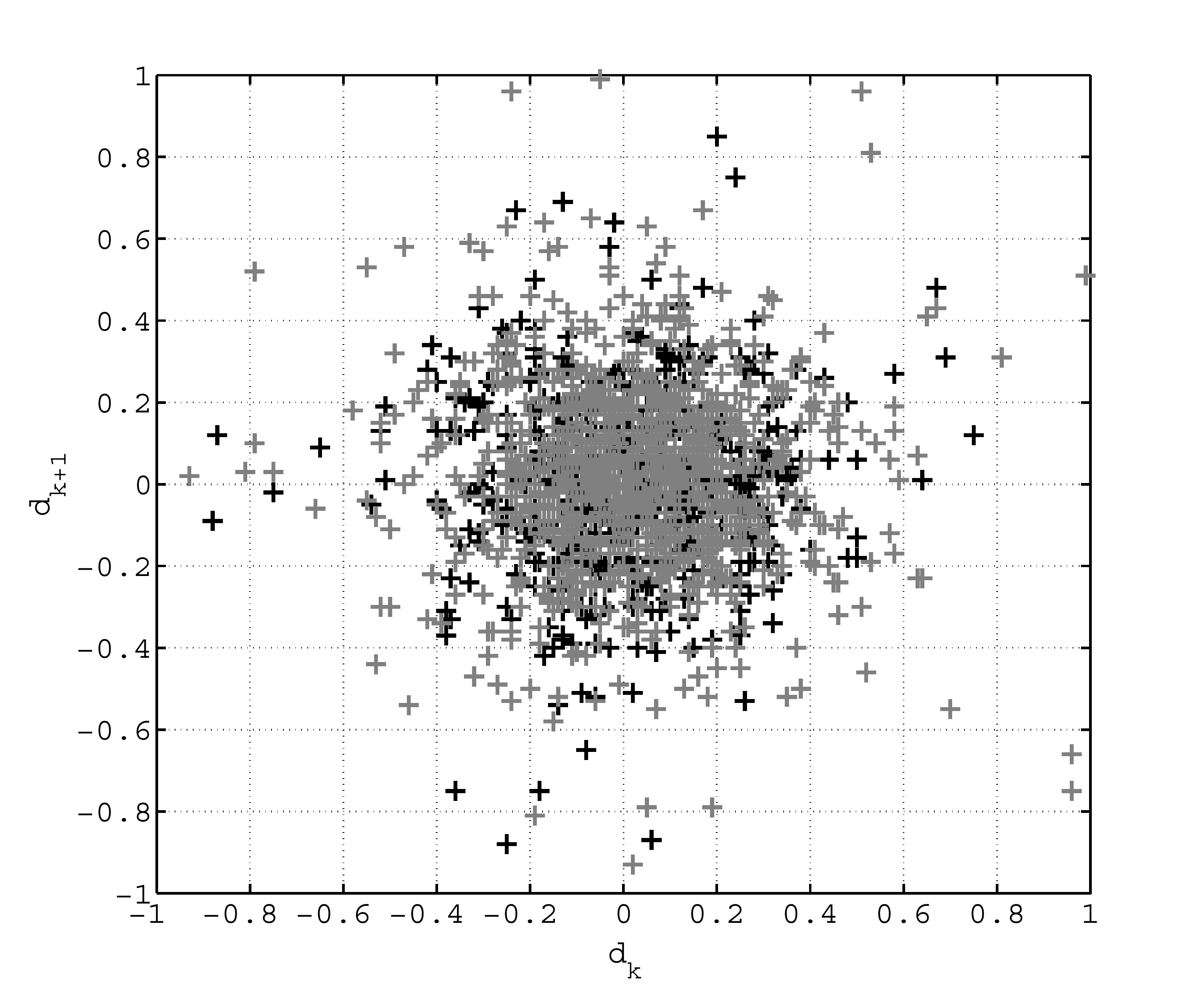}
    }
    \;\;
    \subfloat[Lag-2 plot.]{\label{fig:HST-lagplot2}
        \includegraphics[width=0.31\textwidth]{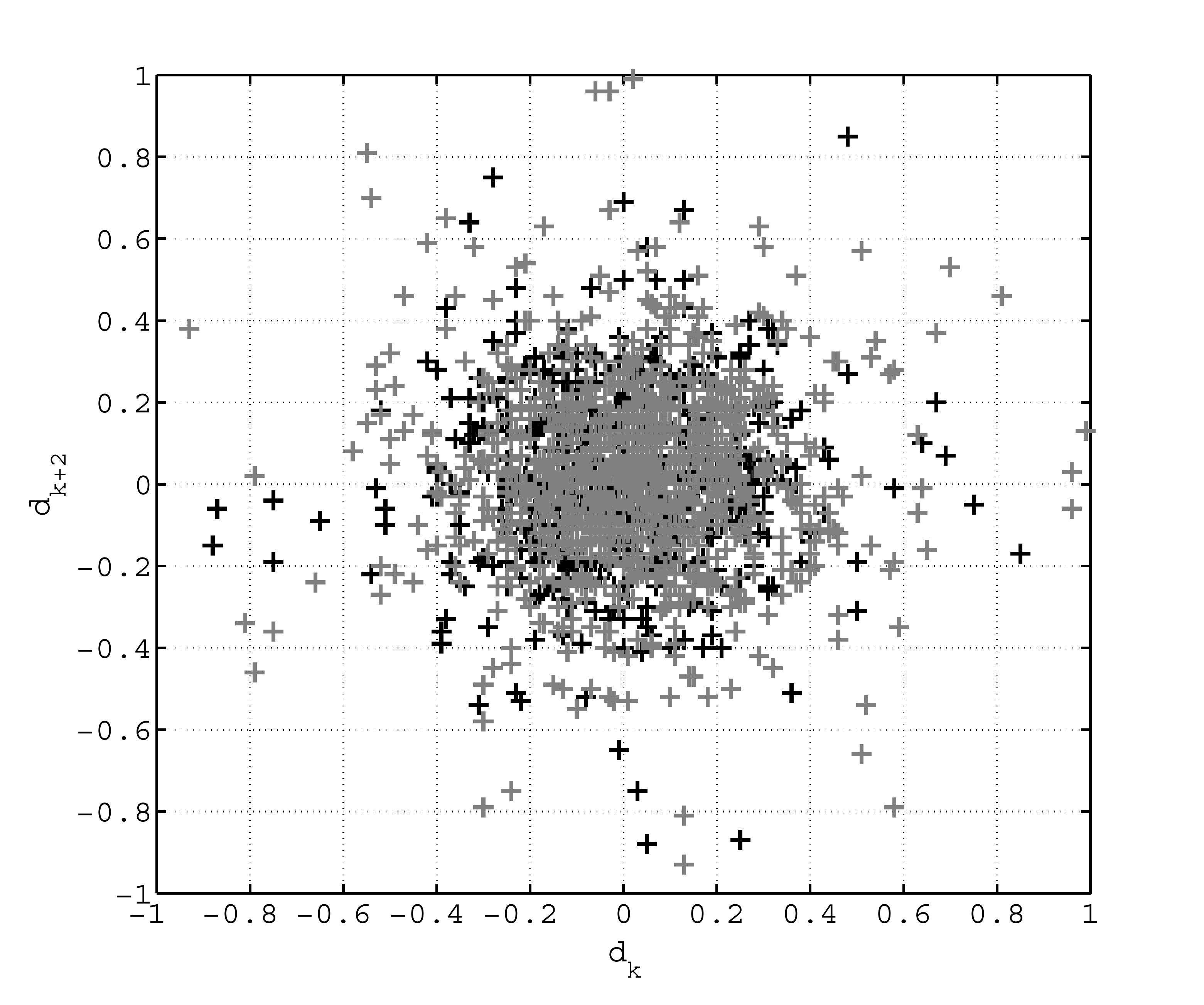}
    }
    \;\;
    \subfloat[Lag-3 plot.]{\label{fig:HST-lagplot3}
        \includegraphics[width=0.31\textwidth]{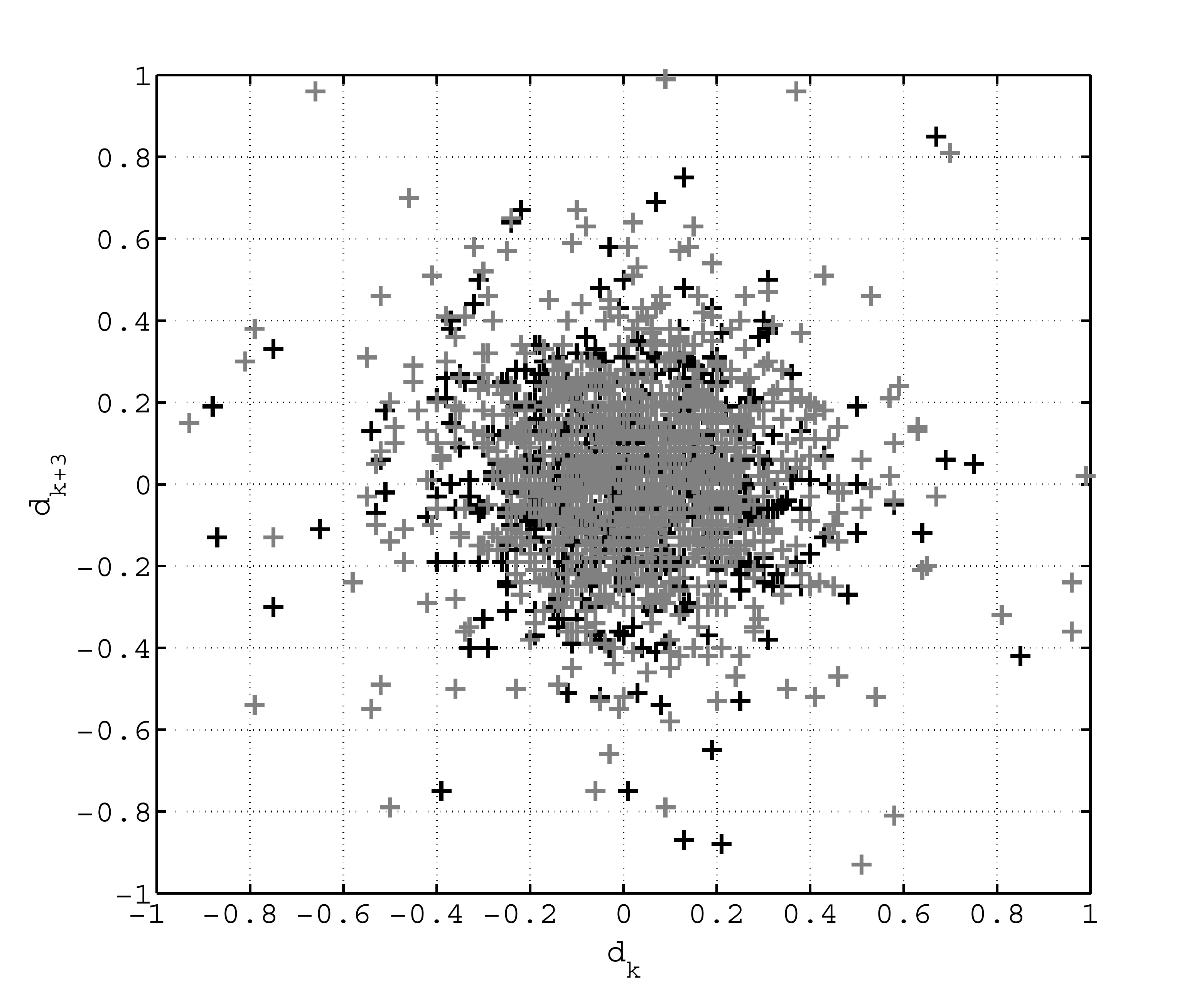}
    }\\
    \subfloat[Lag-11 plot.]{\label{fig:HST-lagplot11}
        \includegraphics[width=0.31\textwidth]{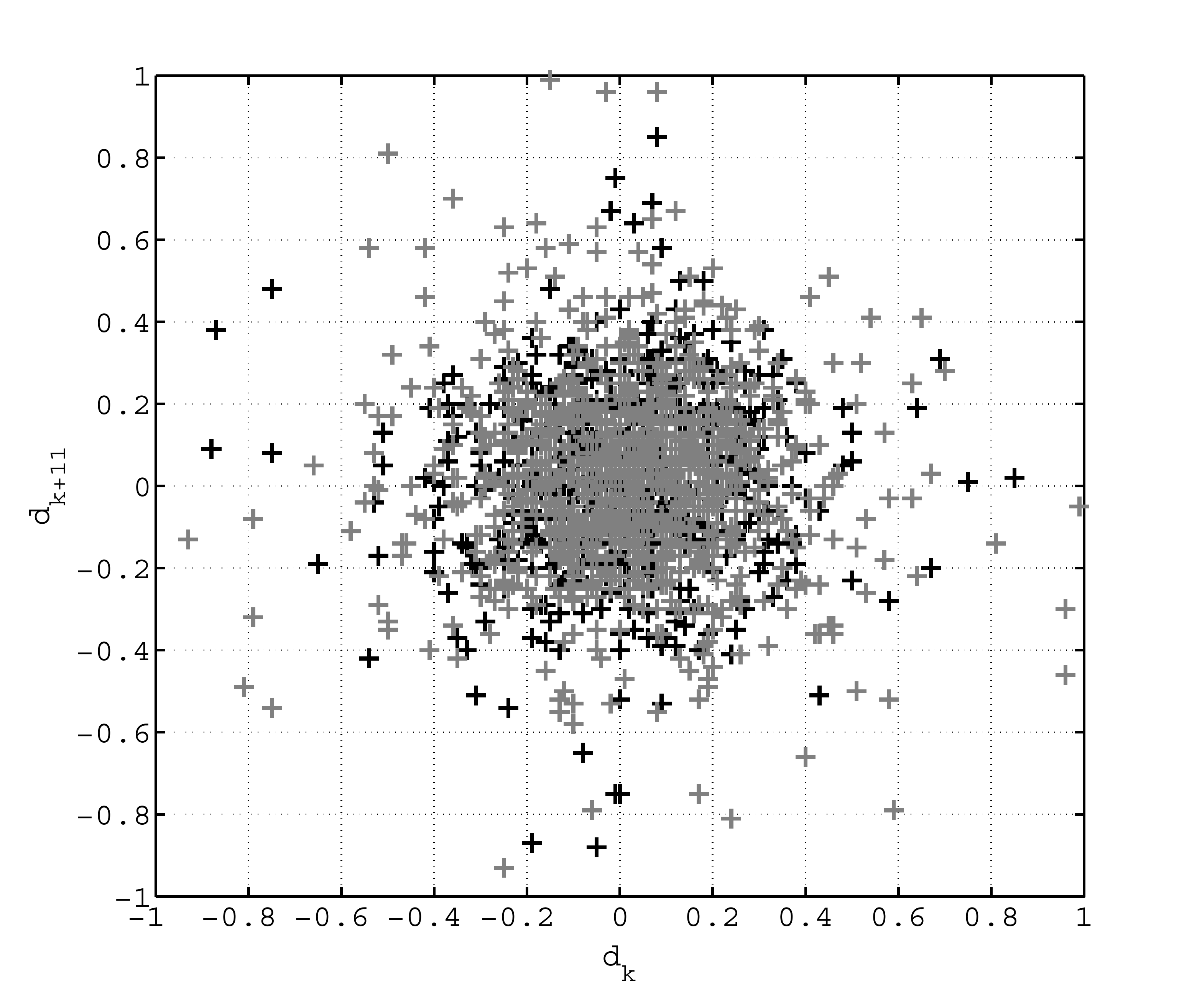}
    }
    \;\;
    \subfloat[Lag-13 plot.]{\label{fig:HST-lagplot13}
        \includegraphics[width=0.31\textwidth]{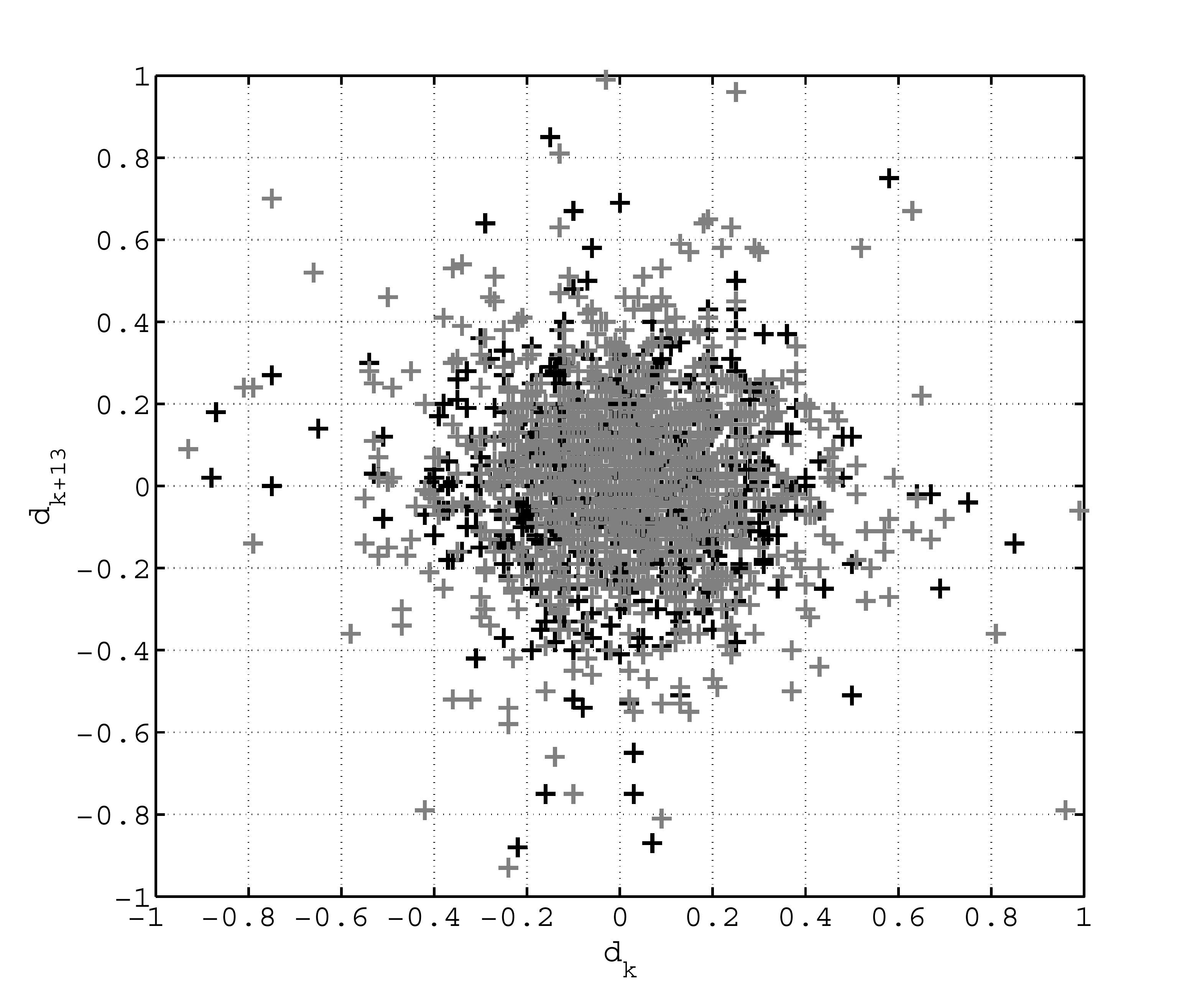}
    }
    \caption{Lag plots (scatter plots) for the HST stock returns.}
    \label{fig:HST-lagplot}
\end{figure*}

To reinforce the ``no-correlation'' conclusion arrived at from Figure~\ref{fig:HST-autocorr}, Figure~\ref{fig:HST-lagplot} provides a selection of lag plots for the data, for lags equal to 1, 2, 3, 11, and 13. According to Figure~\ref{fig:HST-autocorr}, these are the lags at which the data correlation function may be considered statistically significant (with the level of significance being 95\,\%). To clear this out, the scatter plots shown in Figures~\ref{fig:HST-lagplot} are to offer additional insight into the correlation structure of the time series under consideration. As in Figure~\ref{fig:HST-autocorr} above, in Figure~\ref{fig:HST-lagplot} the data are also split into two categories---before the 2003 event and after---and the two categories are distinguished using black color for the first category (before the 2003 event) and gray for the second one (after the 2003 event). The lack of any apparent patters in any of the five lag plots is an indication that the HST stock daily return series exhibits no temporal correlation.

\subsection{Online structural break detection}

We are now in a position to devise the change-point detection methodology of Section~\ref{sec:application} to detect changes in the statistical pattern of the HST returns. To assess the performance of our detection methods, we will measure the detection delay relative to the change-point estimated by the Brodsky--Darkhovsky estimator~\eqref{eq:Brodsky-Darkhovsky-est-def} above. Recall that we are interested in comparing two score-based change-point detection procedures: the CUSUM chart given by~\eqref{eq:Wn-NPCS-def}--\eqref{eq:T-NPCS-def} and the Shiryaev--Roberts (SR) procedure given by~\eqref{eq:Rn-NPSR-def}--\eqref{eq:T-NPSR-def}. Selecting the score function as in~\eqref{eq:score}--\eqref{eq:design_C} for either procedure, we have implemented both detection methods in MATLAB, the well-known scientific computing platform developed by MathWorks, Inc. (see on the Web at~\url{http://www.mathworks.com}). Since the above analysis of the data resulted in the conclusion that the data do follow a Gaussian model (before as well as after the change), to set up the detection threshold of the CUSUM chart and the SR procedure we assumed the Gaussian model with the parameters chosen as estimated in the above analysis. Via a simple Monte Carlo experiment we estimated that setting $A \approx 60$ and $h \approx 0.3$ ensures that the ARL to false alarm of either procedure is approximately $7$ samples,
which is roughly a week, since the timescale is working days.
\begin{figure}[!htb]
    \centering
    \subfloat[By the SR procedure.]{\label{fig:HST-detection:SR}
        \includegraphics[width=0.5\textwidth]{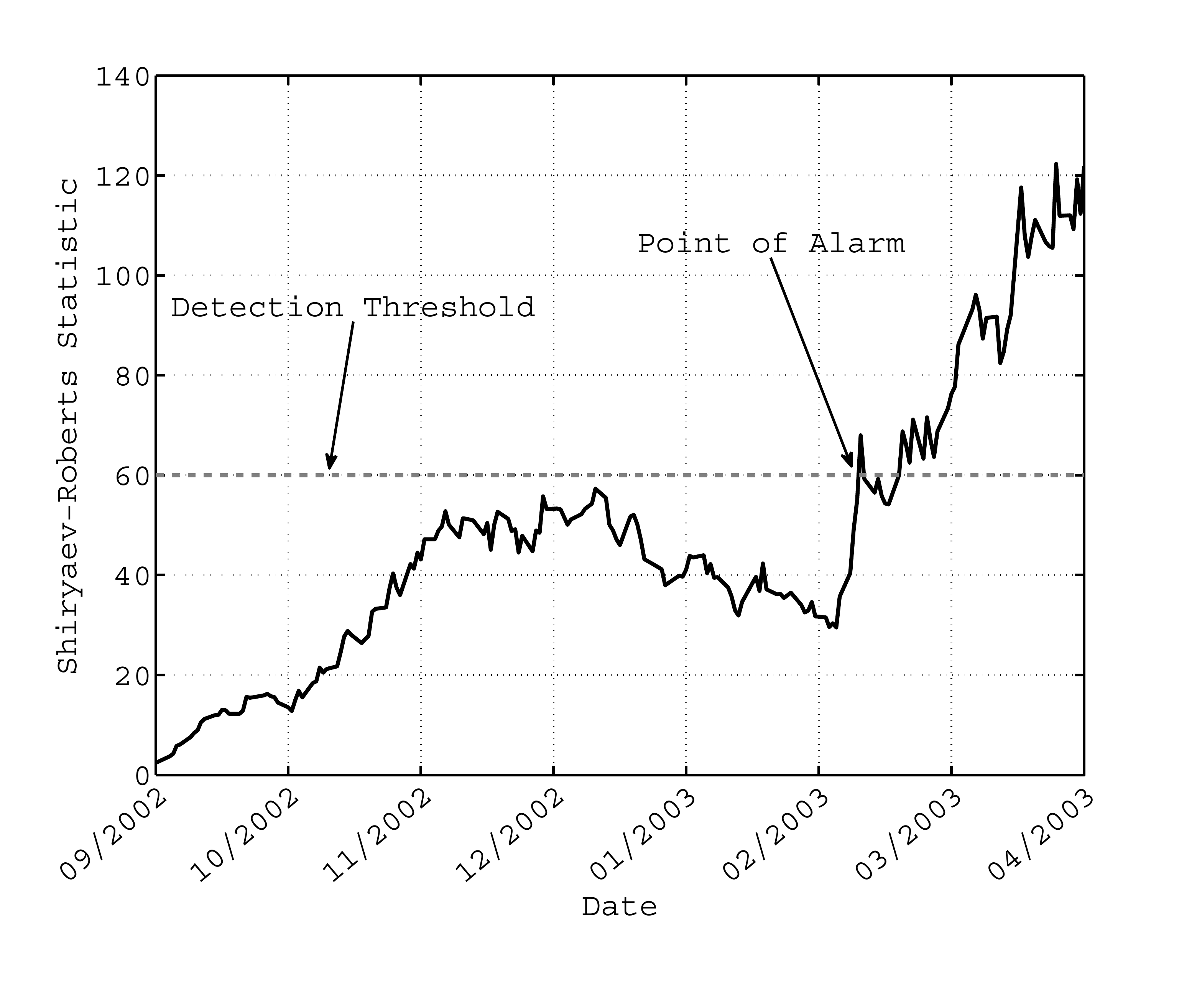}
    }
    \;\;
    \subfloat[By the CUSUM chart.]{\label{fig:HST-detection:CUSUM}
        \includegraphics[width=0.5\textwidth]{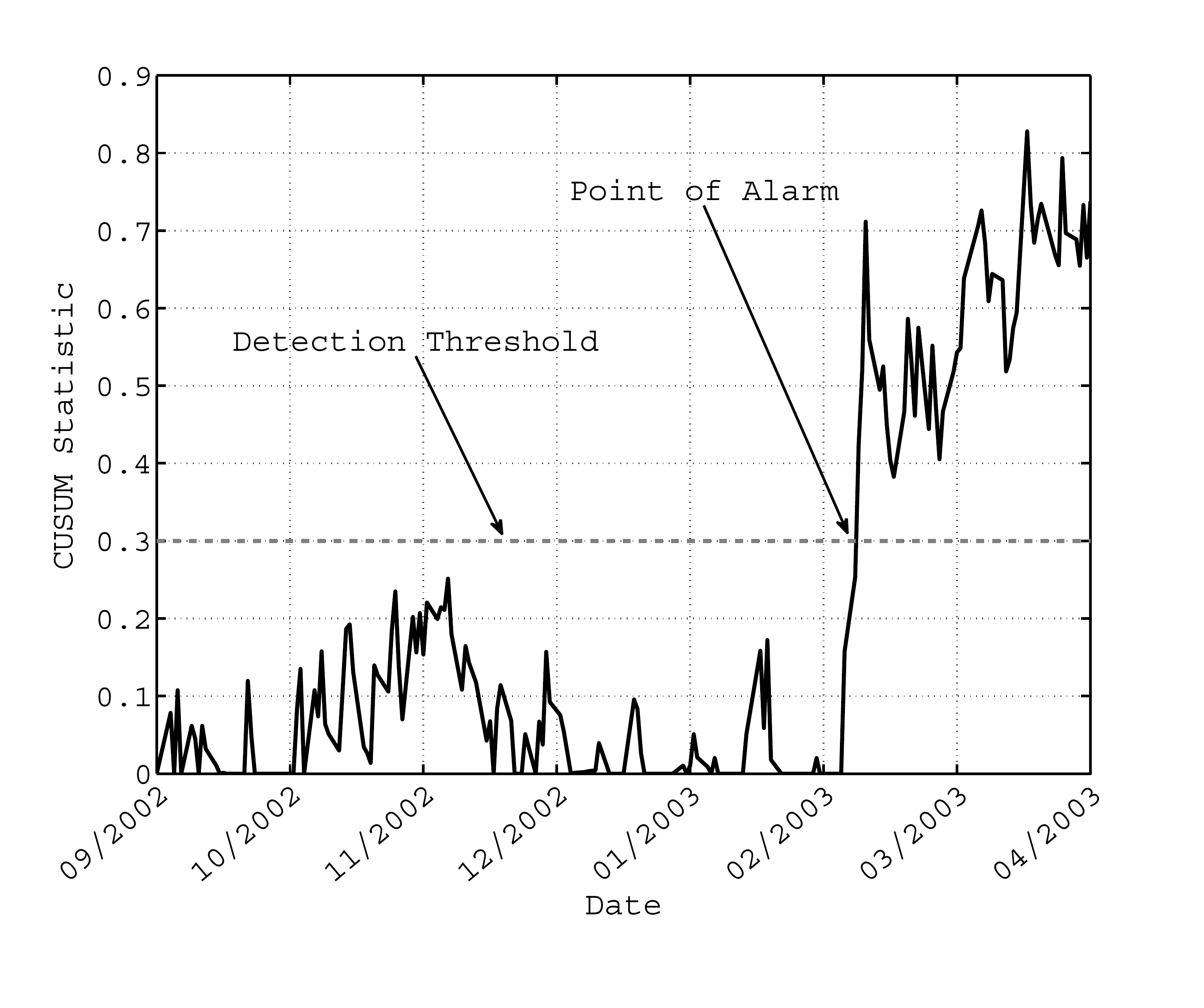}
    }
    \caption{Detection of the 2003 anomaly in the HST stock by the SR and CUSUM procedures.}
    \label{fig:HST-detection}
\end{figure}

The detection process is illustrated in Figure~\ref{fig:HST-detection}. Specifically, Figure~\ref{fig:HST-detection:SR} shows the behavior of the SR statistic in a short time window covering March 13, 2003, i.e., the date at which the HST stock underwent the change we would like to detect. Such a ``zoomed-in'' scale is to better illustrate the dynamics of the detection statistic around the change-point. Figure~\ref{fig:HST-detection:CUSUM} shows the same but for the CUSUM statistic. We see that both procedures successfully detect the onset of the drift (occurring on March 13, 2003), and the detection delays are about one day each.

To draw a line under this section, we would like to remark that the dynamics of the CUSUM statistic is generally more informative than the dynamics of the SR statistic; compare, e.g., Figure~\ref{fig:HST-detection:CUSUM} showing the CUSUM statistic and Figure~\ref{fig:HST-detection:SR} showing the corresponding SR statistic. Specifically, a mere eye examination of the behavior of the CUSUM statistic as a function of time allows not only to see whether the change has occurred or not, but to also estimate the time of its occurrence, i.e., the change-point: it is likely to be somewhere between the time instance the CUSUM statistic last hit zero and the point at which the statistic hit (or went above) the detection threshold (i.e., the point of alarm). Indeed, on the one hand, the change-point is unlikely to be past the point of alarm. On the other hand, as we discussed in the previous section, the CUSUM statistic is effectively a random walk with the ``instantaneous'' LLRs being the increments. Since the LLRs are, on average, negative if no change is in effect, and positive otherwise, the drift of the random walk the CUSUM chart uses for its decision-making is negative before the change and positive after. As a result, the CUSUM statistic effectively estimates zero in the pre-change regime, because zero is its reflection barrier: every time the CUSUM statistic hits zero it resets itself completely ``forgetting'' everything it had previously ``learned'' about the data. This equips the CUSUM chart with a built-in resetting mechanism: if after a sufficiently long period of surveillance the data have given no indication of a change, the CUSUM statistic will likely reset itself (by hitting zero), i.e., it will discard the entire history of observations made up to that point and start completely anew. Hence, the change-point is unlikely to be to the left of the latest point at which the CUSUM statistic visited zero. This intrinsic self-restarting feature is the main reason for the exact minimax optimality (in the sense of Lorden~\cite{Lorden:AMS71}) of the CUSUM chart established in~\cite{Moustakides:AS86,Ritov:AS90}. By contrast, the SR statistic when plotted against time does not offer this kind of convenience of interpretation, for the SR procedure's decision-making mechanism uses entirely different principles. Nevertheless, the SR procedure is exactly multi-cyclic optimal, and the CUSUM chart is not. Therefore, when it comes to monitoring processes that are unlikely to undergo a structural break for a long period of time, so that change-point detection has to be performed in cycles, basing surveillance on the SR procedure might be a better option than going with the CUSUM chart.

\section{Conclusion}
\label{sec:conclusion}

We considered the problem of rapid but reliable anomaly detection in ``live'' financial data. We treated the problem statistically, viz. as that of quickest change-point detection, and proposed an anomaly-detection method that derives from the multi-cyclic (repeated) Shiryaev--Roberts (SR) detection procedure. We decided to go with this largely neglected near-coeval of the celebrated CUSUM and EWMA charts because of the strong multi-cyclic optimality properties that the SR procedure was recently discovered to have under the basic iid change-point detection setup; no such properties are exhibited by either the ``good old'' CUSUM ``inspection scheme'' or the EWMA chart. To handle real-world financial data, the proposed SR-derivative utilizes the information contained in the data in the SR-like Bayesian manner with the likelihood ratio replaced with a change-sensitive score function. This simple idea allowed the proposed procedure to preserve the low computational complexity of its prototype---the original SR procedure. More importantly, we carried out a case study where the proposed procedure was devised to detect an anomaly in a real-world financial time series, and the obtained experimental results indicated that our procedure may have also preserved the great ``false alarm risk''-vs.-``detection speed'' capabilities of the original SR procedure.

\section*{Acknowledgements}
The authors would like to thank the two Guest Editors, Prof.~Shouyang Wang (Institute of Systems Science, Beijing, P.R.~China) and Prof.~Anatoly Zhigljavsky (Cardiff University, Cardiff, Wales, UK), and the Editor-in-Chief, Prof.~Heping Zhang (Yale University, New Haven, Connecticut, USA), for the time and effort they invested to produce this special issue of the Journal. The authors are also personally thankful to Prof.~Zhigljavsky for the invitation to contribute this work to the special issue. The constructive feedback provided by the two anonymous referees is greatly appreciated as well.

The effort of A.S.~Polunchenko was supported, in part, by the Simons Foundation (\url{www.simonsfoundation.org}) via a Collaboration Grant in Mathematics (Award \#\,304574).

A.S.~Polunchenko is also equally indebted to the Office of the Dean of the Harpur College of Arts and Sciences at the State University of New York at Binghamton for the support provided through the Dean's Research Semester Award for Junior Faculty granted for the Fall semester of 2014.

The work of A.~Pepelyshev was partly supported by the St.~Petersburg State University, Russia, under project \#\,6.38.435.2015.


\bibliographystyle{imsart-number}      
\bibliography{main,finance,ssa,spc,hst}

\begin{thebibliography}{59}

\bibitem{Basseville+Nikiforov:Book93}
\begin{bbook}[author]
\bauthor{\bsnm{Basseville},~\bfnm{Mich\`{e}le}\binits{M.}} \AND
  \bauthor{\bsnm{Nikiforov},~\bfnm{Igor~V.}\binits{I.~V.}}
(\byear{1993}).
\btitle{Detection of Abrupt Changes: Theory and Application}.
\bpublisher{Prentice Hall}, \baddress{Englewood Cliffs, NJ}.
\end{bbook}
\endbibitem

\bibitem{Brodsky+Darkhovsky:Book1993}
\begin{bbook}[author]
\bauthor{\bsnm{Brodsky},~\bfnm{Boris~E.}\binits{B.~E.}} \AND
  \bauthor{\bsnm{Darkhovsky},~\bfnm{Boris~S.}\binits{B.~S.}}
(\byear{1993}).
\btitle{Nonparametric Methods in Change-Point Problems}.
\bseries{Mathematics and Its Aplications}
\bvolume{243}.
\bpublisher{Kluwer Academic Publishers}, \baddress{Norwell, MA}.
\end{bbook}
\endbibitem

\bibitem{Brodsky+Darkhovsky:Book2000}
\begin{bbook}[author]
\bauthor{\bsnm{Brodsky},~\bfnm{Boris~E.}\binits{B.~E.}} \AND
  \bauthor{\bsnm{Darkhovsky},~\bfnm{Boris~S.}\binits{B.~S.}}
(\byear{2000}).
\btitle{Non-Parametric Statistical Diagnosis: Problems and Methods}.
\bseries{Mathematics and Its Aplications}
\bvolume{509}.
\bpublisher{Kluwer Academic Publishers}, \baddress{Norwell, MA}.
\end{bbook}
\endbibitem

\bibitem{Dragalin:PSIM94}
\begin{bincollection}[author]
\bauthor{\bsnm{Dragalin},~\bfnm{Vladimir~P.}\binits{V.~P.}}
(\byear{1994}).
\btitle{Optimality of a generalized {CUSUM} procedure in quickest detection
  problem}.
In \bbooktitle{Proceedings of the {S}teklov Institute of Mathematics:
  Statistics and Control of Random Processes},
\bvolume{202}
\bpages{107--120}.
\bpublisher{American Mathematical Society}, \baddress{Providence, RI}.
\end{bincollection}
\endbibitem

\bibitem{Du+etal:CommStat2015}
\begin{barticle}[author]
\bauthor{\bsnm{Du},~\bfnm{Wenyu}\binits{W.}},
  \bauthor{\bsnm{Polunchenko},~\bfnm{Aleksey~S.}\binits{A.~S.}} \AND
  \bauthor{\bsnm{Sokolov},~\bfnm{Grigory}\binits{G.}}
(\byear{2015}).
\btitle{On Robustness of the {S}hiryaev--{R}oberts Change-Point Detection
  Procedure under Parameter Misspecification in the Post-Change Distribution}.
\bjournal{Communications in Statistics---Simulation and Computation}.
\bnote{(in press). Available online
  at:~\url{http://www.tandfonline.com/doi/full/10.1080/03610918.2015.1039131}}.
\bdoi{10.1080/03610918.2015.1039131}
\end{barticle}
\endbibitem

\bibitem{Ergashev:EconWPA2004}
\begin{bmisc}[author]
\bauthor{\bsnm{Ergashev},~\bfnm{Bakhodir~A.}\binits{B.~A.}}
(\byear{2004}).
\btitle{Sequential Detection of {US} Business Cycle Turning Points:
  Performances of {S}hiryayev--{R}oberts, {CUSUM} and {EWMA} Procedures}.
\bnote{Available online at the Economics Working Paper Archive
  (EconWPA):~\url{https://ideas.repec.org/p/wpa/wuwpem/0402001.html}}.
\end{bmisc}
\endbibitem

\bibitem{Frisen:Book2008}
\begin{bbook}[author]
\bauthor{\bsnm{Fris\'en},~\bfnm{Marianne}\binits{M.}}
(\byear{2008}).
\btitle{Financial Surveillance}.
\bpublisher{John Wiley \& Sons, Inc.}, \baddress{Hoboken, NJ}.
\end{bbook}
\endbibitem

\bibitem{Frisen:SA2009}
\begin{barticle}[author]
\bauthor{\bsnm{Fris\'en},~\bfnm{Marianne}\binits{M.}}
(\byear{2009}).
\btitle{Optimal Sequential Surveillance for Finance, Public Health, and Other
  Areas}.
\bjournal{Sequential Analysis}
\bvolume{28}
\bpages{310--337}.
\bdoi{10.1080/07474940903041605}
\end{barticle}
\endbibitem

\bibitem{Girschick+Rubin:AMS52}
\begin{barticle}[author]
\bauthor{\bsnm{Girschick},~\bfnm{Meyer~Abraham}\binits{M.~A.}} \AND
  \bauthor{\bsnm{Rubin},~\bfnm{Herman}\binits{H.}}
(\byear{1952}).
\btitle{A {B}ayes approach to a quality control model}.
\bjournal{Annals of Mathematical Statistics}
\bvolume{23}
\bpages{114--125}.
\bdoi{10.1214/aoms/1177729489}
\end{barticle}
\endbibitem

\bibitem{Golyandina+etal:Book2001}
\begin{bbook}[author]
\bauthor{\bsnm{Golyandina},~\bfnm{Nina}\binits{N.}},
  \bauthor{\bsnm{Nekrutkin},~\bfnm{Vladimir}\binits{V.}} \AND
  \bauthor{\bsnm{Zhigljavsky},~\bfnm{Anatoly~A}\binits{A.~A.}}
(\byear{2001}).
\btitle{Analysis of Time Series Structure: {SSA} and related techniques}.
\bseries{Monographs on Statistics and Applied Probability}
\bvolume{90}.
\bpublisher{Chapman \& Hall/CRC}, \baddress{London, UK}.
\end{bbook}
\endbibitem

\bibitem{Golyandina+Zhigljavsky:Book2013}
\begin{bbook}[author]
\bauthor{\bsnm{Golyandina},~\bfnm{Nina}\binits{N.}} \AND
  \bauthor{\bsnm{Zhigljavsky},~\bfnm{Anatoly}\binits{A.}}
(\byear{2013}).
\btitle{Singular Spectrum Analysis for Time Series}.
\bseries{Springer Briefs in Statistics}.
\bpublisher{Springer}.
\end{bbook}
\endbibitem

\bibitem{Gordon+Pollak:AS1994}
\begin{barticle}[author]
\bauthor{\bsnm{Gordon},~\bfnm{Louis}\binits{L.}} \AND
  \bauthor{\bsnm{Pollak},~\bfnm{Moshe}\binits{M.}}
(\byear{1994}).
\btitle{An Efficient Sequential Nonparametric Scheme for Detecting a Change of
  Distribution}.
\bjournal{Annals of Statistics}
\bvolume{22}
\bpages{763--804}.
\bdoi{10.1214/aos/1176325495}
\end{barticle}
\endbibitem

\bibitem{Gordon+Pollak:AS1995}
\begin{barticle}[author]
\bauthor{\bsnm{Gordon},~\bfnm{Louis}\binits{L.}} \AND
  \bauthor{\bsnm{Pollak},~\bfnm{Moshe}\binits{M.}}
(\byear{1995}).
\btitle{A Robust Surveillance Scheme for Stochastically Ordered Alternatives}.
\bjournal{Annals of Statistics}
\bvolume{23}
\bpages{1350--1375}.
\bdoi{10.1214/aos/1176324712}
\end{barticle}
\endbibitem

\bibitem{HST:Form10K-2001}
\begin{bbooklet}[author]
\bauthor{\bsnm{{Host Hotels \& Resorts, Inc. }}}
(\byear{2001}).
\btitle{{US} {S}ecurities and {E}xchange {C}ommission, {F}orm {10-K}, {T}ax
  {Y}ear 2001}.
\end{bbooklet}
\endbibitem

\bibitem{HST:AR-2002}
\begin{bbooklet}[author]
\bauthor{\bsnm{{Host Hotels \& Resorts, Inc. }}}
(\byear{2002}).
\btitle{Annual Report 2002}.
\end{bbooklet}
\endbibitem

\bibitem{HST:AR-2003}
\begin{bbooklet}[author]
\bauthor{\bsnm{{Host Hotels \& Resorts, Inc. }}}
(\byear{2003}).
\btitle{Annual Report 2003}.
\end{bbooklet}
\endbibitem

\bibitem{HST:Form10K-2014}
\begin{bbooklet}[author]
\bauthor{\bsnm{{Host Hotels \& Resorts, Inc. }}}
(\byear{2014}).
\btitle{{US} {S}ecurities and {E}xchange {C}ommission, {F}orm {10-K}, {T}ax
  {Y}ear 2014}.
\end{bbooklet}
\endbibitem

\bibitem{Lai:JRSS95}
\begin{barticle}[author]
\bauthor{\bsnm{Lai},~\bfnm{Tze~Leung}\binits{T.~L.}}
(\byear{1995}).
\btitle{Sequential changepoint detection in quality control and dynamical
  systems (with discussion)}.
\bjournal{Journal of the Royal Statistical Society. Series B. Methodological}
\bvolume{57}
\bpages{613--658}.
\end{barticle}
\endbibitem

\bibitem{Lai:IEEE-IT1998}
\begin{barticle}[author]
\bauthor{\bsnm{Lai},~\bfnm{Tze~Leung}\binits{T.~L.}}
(\byear{1998}).
\btitle{Information bounds and quick detection of parameter changes in
  stochastic systems}.
\bjournal{{IEEE} Transactions on Information Theory}
\bvolume{44}
\bpages{2917--2929}.
\bdoi{10.1109/18.737522}
\end{barticle}
\endbibitem

\bibitem{Lai+Xing:Book2015}
\begin{bbook}[author]
\bauthor{\bsnm{Lai},~\bfnm{Tze~Leung}\binits{T.~L.}} \AND
  \bauthor{\bsnm{Xing},~\bfnm{Haipeng}\binits{H.}}
(\byear{2015}).
\btitle{Active Risk Management: Financial Models and Statistical Methods}.
\bseries{Chapman and Hall/CRC Financial Mathematics Series}.
\bpublisher{Chapman \& Hall/CRC Press}, \baddress{Boca Raton, FL}.
\end{bbook}
\endbibitem

\bibitem{Lorden:AMS71}
\begin{barticle}[author]
\bauthor{\bsnm{Lorden},~\bfnm{Gary}\binits{G.}}
(\byear{1971}).
\btitle{Procedures for reacting to a change in distribution}.
\bjournal{Annals of Mathematical Statistics}
\bvolume{42}
\bpages{1897--1908}.
\bdoi{10.1214/aoms/1177693055}
\end{barticle}
\endbibitem

\bibitem{McDonald:NRL1990}
\begin{barticle}[author]
\bauthor{\bsnm{McDonald},~\bfnm{David}\binits{D.}}
(\byear{1990}).
\btitle{A {C}USUM Procedure Based on Sequential Ranks}.
\bjournal{Journal of Naval Research}
\bvolume{37}
\bpages{627--646}.
\end{barticle}
\endbibitem

\bibitem{Moskvina+Zhigljavsky:CommStat2003}
\begin{barticle}[author]
\bauthor{\bsnm{Moskvina},~\bfnm{Valentina}\binits{V.}} \AND
  \bauthor{\bsnm{Zhigljavsky},~\bfnm{Anatoly}\binits{A.}}
(\byear{2003}).
\btitle{An Algorithm Based on Singular Spectrum Analysis for Change-Point
  Detection}.
\bjournal{Communications in Statistics---Simulation and Computation}
\bvolume{32}
\bpages{319--352}.
\bdoi{10.1081/SAC-120017494}
\end{barticle}
\endbibitem

\bibitem{Moustakides:AS86}
\begin{barticle}[author]
\bauthor{\bsnm{Moustakides},~\bfnm{George~V.}\binits{G.~V.}}
(\byear{1986}).
\btitle{Optimal stopping times for detecting changes in distributions}.
\bjournal{Annals of Statistics}
\bvolume{14}
\bpages{1379--1387}.
\end{barticle}
\endbibitem

\bibitem{Page:B54}
\begin{barticle}[author]
\bauthor{\bsnm{Page},~\bfnm{Ewan~S.}\binits{E.~S.}}
(\byear{1954}).
\btitle{Continuous Inspection Schemes}.
\bjournal{Biometrika}
\bvolume{41}
\bpages{100--115}.
\bdoi{10.1093/biomet/41.1-2.100}
\end{barticle}
\endbibitem

\bibitem{Pollak:AS85}
\begin{barticle}[author]
\bauthor{\bsnm{Pollak},~\bfnm{Moshe}\binits{M.}}
(\byear{1985}).
\btitle{Optimal detection of a change in distribution}.
\bjournal{Annals of Statistics}
\bvolume{13}
\bpages{206--227}.
\bdoi{10.1214/aos/1176346587}
\end{barticle}
\endbibitem

\bibitem{Pollak:AS87}
\begin{barticle}[author]
\bauthor{\bsnm{Pollak},~\bfnm{Moshe}\binits{M.}}
(\byear{1987}).
\btitle{Average run lengths of an optimal method of detecting a change in
  distribution}.
\bjournal{Annals of Statistics}
\bvolume{15}
\bpages{749--779}.
\bdoi{10.1214/aos/1176350373}
\end{barticle}
\endbibitem

\bibitem{Pollak:SA2010}
\begin{barticle}[author]
\bauthor{\bsnm{Pollak},~\bfnm{Moshe}\binits{M.}}
(\byear{2010}).
\btitle{A Robust Changepoint Detection Method}.
\bjournal{Sequential Analysis}
\bvolume{29}
\bpages{146--161}.
\bdoi{10.1080/07474941003741029}
\end{barticle}
\endbibitem

\bibitem{Pollak+Krieger:SA2013}
\begin{barticle}[author]
\bauthor{\bsnm{Pollak},~\bfnm{Moshe}\binits{M.}} \AND
  \bauthor{\bsnm{Krieger},~\bfnm{Abba~M.}\binits{A.~M.}}
(\byear{2013}).
\btitle{Shewhart Revisited}.
\bjournal{Sequential Analysis}
\bvolume{32}
\bpages{230--242}.
\bdoi{10.1080/07474946.2013.774621}
\end{barticle}
\endbibitem

\bibitem{Pollak+Tartakovsky:ISITA2008}
\begin{binproceedings}[author]
\bauthor{\bsnm{Pollak},~\bfnm{Moshe}\binits{M.}} \AND
  \bauthor{\bsnm{Tartakovsky},~\bfnm{Alexander~G.}\binits{A.~G.}}
(\byear{2008}).
\btitle{Exact Optimality of the {S}hiryaev--{R}oberts Procedure for Detecting
  Changes in Distributions}.
In \bbooktitle{Proceedings of the 2008 International Symposium on Information
  Theory and Its Applications}
\bpages{1--6}.
\bdoi{10.1109/ISITA.2008.4895424}
\end{binproceedings}
\endbibitem

\bibitem{Pollak+Tartakovsky:SS09}
\begin{barticle}[author]
\bauthor{\bsnm{Pollak},~\bfnm{Moshe}\binits{M.}} \AND
  \bauthor{\bsnm{Tartakovsky},~\bfnm{Alexander~G.}\binits{A.~G.}}
(\byear{2009}).
\btitle{Optimality Properties of the {S}hiryaev--{R}oberts procedure}.
\bjournal{Statistica Sinica}
\bvolume{19}
\bpages{1729--1739}.
\end{barticle}
\endbibitem

\bibitem{Polunchenko+etal:JSM2013}
\begin{binproceedings}[author]
\bauthor{\bsnm{Polunchenko},~\bfnm{Aleksey~S.}\binits{A.~S.}},
  \bauthor{\bsnm{Sokolov},~\bfnm{Grigory}\binits{G.}} \AND
  \bauthor{\bsnm{Du},~\bfnm{Wenyu}\binits{W.}}
(\byear{2013}).
\btitle{Quickest Change-Point Detection: A Bird's Eye View}.
In \bbooktitle{Proceedings of the 2013 Joint Statistical Meetings}.
\end{binproceedings}
\endbibitem

\bibitem{Polunchenko+Tartakovsky:AS10}
\begin{barticle}[author]
\bauthor{\bsnm{Polunchenko},~\bfnm{Aleksey~S.}\binits{A.~S.}} \AND
  \bauthor{\bsnm{Tartakovsky},~\bfnm{Alexander~G.}\binits{A.~G.}}
(\byear{2010}).
\btitle{On optimality of the {S}hiryaev--{R}oberts procedure for detecting a
  change in distribution}.
\bjournal{Annals of Statistics}
\bvolume{38}
\bpages{3445--3457}.
\bdoi{10.1214/09-AOS775}
\end{barticle}
\endbibitem

\bibitem{Polunchenko+Tartakovsky:MCAP2012}
\begin{barticle}[author]
\bauthor{\bsnm{Polunchenko},~\bfnm{Aleksey~S.}\binits{A.~S.}} \AND
  \bauthor{\bsnm{Tartakovsky},~\bfnm{Alexander~G.}\binits{A.~G.}}
(\byear{2012}).
\btitle{State-of-the-Art in Sequential Change-Point Detection}.
\bjournal{Methodology and Computing in Applied Probability}
\bvolume{14}
\bpages{649--684}.
\bdoi{10.1007/s11009-011-9256-5}
\end{barticle}
\endbibitem

\bibitem{Polunchenko+etal:SA2012}
\begin{barticle}[author]
\bauthor{\bsnm{Polunchenko},~\bfnm{Aleksey~S.}\binits{A.~S.}},
  \bauthor{\bsnm{Tartakovsky},~\bfnm{Alexander~G.}\binits{A.~G.}} \AND
  \bauthor{\bsnm{Mukhopadhyay},~\bfnm{Nitis}\binits{N.}}
(\byear{2012}).
\btitle{Nearly Optimal Change-Point Detection with an Application to
  Cybersecurity}.
\bjournal{Sequential Analysis}
\bvolume{31}
\bpages{409--435}.
\bdoi{10.1080/07474946.2012.694351}
\end{barticle}
\endbibitem

\bibitem{Poor+Hadjiliadis:Book09}
\begin{bbook}[author]
\bauthor{\bsnm{Poor},~\bfnm{H.~Vincent}\binits{H.~V.}} \AND
  \bauthor{\bsnm{Hadjiliadis},~\bfnm{Olympia}\binits{O.}}
(\byear{2009}).
\btitle{Quickest Detection}.
\bpublisher{Cambridge University Press}, \baddress{New York, NY}.
\end{bbook}
\endbibitem

\bibitem{Ritov:AS90}
\begin{barticle}[author]
\bauthor{\bsnm{Ritov},~\bfnm{Ya'acov}\binits{Y.}}
(\byear{1990}).
\btitle{Decision theoretic optimality of the {CUSUM} procedure}.
\bjournal{Annals of Statistics}
\bvolume{18}
\bpages{1464--1469}.
\end{barticle}
\endbibitem

\bibitem{Roberts:T59}
\begin{barticle}[author]
\bauthor{\bsnm{Roberts},~\bfnm{S.~W.}\binits{S.~W.}}
(\byear{1959}).
\btitle{Control chart tests based on geometric moving averages}.
\bjournal{Technometrics}
\bvolume{1}
\bpages{239--250}.
\bdoi{10.1080/00401706.1959.10489860}
\end{barticle}
\endbibitem

\bibitem{Roberts:T66}
\begin{barticle}[author]
\bauthor{\bsnm{Roberts},~\bfnm{S.~W.}\binits{S.~W.}}
(\byear{1966}).
\btitle{A comparison of some control chart procedures}.
\bjournal{Technometrics}
\bvolume{8}
\bpages{411--430}.
\end{barticle}
\endbibitem

\bibitem{Shewhart:JASA1925}
\begin{barticle}[author]
\bauthor{\bsnm{Shewhart},~\bfnm{Walter~A.}\binits{W.~A.}}
(\byear{1925}).
\btitle{The application of statistics as an aid in maintaining quality of a
  manufactured product}.
\bjournal{Journal of the American Statistical Association}
\bvolume{20}
\bpages{546--548}.
\bdoi{10.2307/2277170}
\end{barticle}
\endbibitem

\bibitem{Shewhart:Book1931}
\begin{bbook}[author]
\bauthor{\bsnm{Shewhart},~\bfnm{Walter~Andrew}\binits{W.~A.}}
(\byear{1931}).
\btitle{Economic Control of Quality of Manufactured Product}.
\bseries{Bell Telephone Laboratories series}.
\bpublisher{D. Van Nostrand Company, Inc.}, \baddress{Princeton, NJ}.
\end{bbook}
\endbibitem

\bibitem{Shiryaev:SMD61}
\begin{barticle}[author]
\bauthor{\bsnm{Shiryaev},~\bfnm{Albert~N.}\binits{A.~N.}}
(\byear{1961}).
\btitle{The problem of the most rapid detection of a disturbance in a
  stationary process}.
\bjournal{Soviet Mathematics--Doklady}
\bvolume{2}
\bpages{795--799}.
\bnote{Translation from Dokl. Akad. Nauk SSSR 138:1039--1042, 1961}.
\end{barticle}
\endbibitem

\bibitem{Shiryaev:TPA63}
\begin{barticle}[author]
\bauthor{\bsnm{Shiryaev},~\bfnm{Albert~N.}\binits{A.~N.}}
(\byear{1963}).
\btitle{On optimum methods in quickest detection problems}.
\bjournal{Theory of Probability and Its Applications}
\bvolume{8}
\bpages{22--46}.
\bdoi{10.1137/1108002}
\end{barticle}
\endbibitem

\bibitem{Shiryaev:Book78}
\begin{bbook}[author]
\bauthor{\bsnm{Shiryaev},~\bfnm{Albert~N.}\binits{A.~N.}}
(\byear{1978}).
\btitle{Optimal Stopping Rules}.
\bpublisher{Springer-Verlag}, \baddress{New York, NY}.
\end{bbook}
\endbibitem

\bibitem{Shiryaev+Zryumov:Khabanov2010}
\begin{bincollection}[author]
\bauthor{\bsnm{Shiryaev},~\bfnm{Albert~N.}\binits{A.~N.}} \AND
  \bauthor{\bsnm{Zryumov},~\bfnm{Pavel~Y.}\binits{P.~Y.}}
(\byear{2010}).
\btitle{On the Linear and Nonlinear Generalized {B}ayesian Disorder Problem
  (Discrete Time Case)}.
In \bbooktitle{Optimality and Risk---Modern Trends in Mathematical Finance}
(\beditor{\bfnm{Freddy}\binits{F.}~\bsnm{Delbaen}},
  \beditor{\bfnm{Mikl{\'o}s}\binits{M.}~\bsnm{R{\'a}sonyi}} \AND
  \beditor{\bfnm{Christophe}\binits{C.}~\bsnm{Stricker}}, eds.)
\bpages{227--236}.
\bpublisher{Springer Berlin Heidelberg}.
\bdoi{10.1007/978-3-642-02608-9_12}
\end{bincollection}
\endbibitem

\bibitem{Siegmund:Book85}
\begin{bbook}[author]
\bauthor{\bsnm{Siegmund},~\bfnm{David}\binits{D.}}
(\byear{1985}).
\btitle{Sequential Analysis: Tests and Confidence Intervals}.
\bseries{Springer Series in Statistics}.
\bpublisher{Springer-Verlag}, \baddress{New York, NY}.
\bdoi{10.1007/978-1-4757-1862-1}
\end{bbook}
\endbibitem

\bibitem{Tartakovsky+etal:Book2014}
\begin{bbook}[author]
\bauthor{\bsnm{Tartakovsky},~\bfnm{Alexander}\binits{A.}},
  \bauthor{\bsnm{Nikiforov},~\bfnm{Igor}\binits{I.}} \AND
  \bauthor{\bsnm{Basseville},~\bfnm{Mich\`{e}le}\binits{M.}}
(\byear{2014}).
\btitle{Sequential Analysis: Hypothesis Testing and Changepoint Detection}.
\bseries{Monographs on Statistics and Applied Probability}
\bvolume{166}.
\bpublisher{CRC Press}, \baddress{Boca Raton, FL}.
\end{bbook}
\endbibitem

\bibitem{Tartakovsky:IEEE-CDC05}
\begin{binproceedings}[author]
\bauthor{\bsnm{Tartakovsky},~\bfnm{Alexander~G.}\binits{A.~G.}}
(\byear{2005}).
\btitle{Asymptotic performance of a multichart {CUSUM} test under false alarm
  probability constraint}.
In \bbooktitle{{IEEE} Conference on Decision and Control}
\bvolume{44}
\bpages{320-325}.
\bdoi{10.1109/CDC.2005.1582175}
\end{binproceedings}
\endbibitem

\bibitem{Tartakovsky+Moustakides:SA10}
\begin{barticle}[author]
\bauthor{\bsnm{Tartakovsky},~\bfnm{Alexander~G.}\binits{A.~G.}} \AND
  \bauthor{\bsnm{Moustakides},~\bfnm{George~V.}\binits{G.~V.}}
(\byear{2010}).
\btitle{State-of-the-Art in {B}ayesian Changepoint Detection}.
\bjournal{Sequential Analysis}
\bvolume{29}
\bpages{125--145}.
\bdoi{10.1080/07474941003740997}
\end{barticle}
\endbibitem

\bibitem{Tartakovsky+etal:TPA2012}
\begin{barticle}[author]
\bauthor{\bsnm{Tartakovsky},~\bfnm{Alexander~G.}\binits{A.~G.}},
  \bauthor{\bsnm{Pollak},~\bfnm{Moshe}\binits{M.}} \AND
  \bauthor{\bsnm{Polunchenko},~\bfnm{Aleksey~S.}\binits{A.~S.}}
(\byear{2012}).
\btitle{Third-order asymptotic optimality of the {G}eneralized
  {S}hiryaev--{R}oberts changepoint detection procedures}.
\bjournal{Theory of Probability and Its Applications}
\bvolume{56}
\bpages{457--484}.
\bdoi{10.1137/S0040585X97985534}
\end{barticle}
\endbibitem

\bibitem{Tartakovsky+Polunchenko:IWAP10}
\begin{binproceedings}[author]
\bauthor{\bsnm{Tartakovsky},~\bfnm{Alexander~G.}\binits{A.~G.}} \AND
  \bauthor{\bsnm{Polunchenko},~\bfnm{Aleksey~S.}\binits{A.~S.}}
(\byear{2010}).
\btitle{Minimax Optimality of the {S}hiryaev--{R}oberts Procedure}.
In \bbooktitle{Proceedings of the 5th {I}nternational {W}orkshop on {A}pplied
  {P}robability}.
\end{binproceedings}
\endbibitem

\bibitem{Tartakovsy+etal:IEEE-JSTSP2013}
\begin{barticle}[author]
\bauthor{\bsnm{Tartakovsky},~\bfnm{Alexander~G.}\binits{A.~G.}},
  \bauthor{\bsnm{Polunchenko},~\bfnm{Aleksey~S.}\binits{A.~S.}} \AND
  \bauthor{\bsnm{Sokolov},~\bfnm{Grigory}\binits{G.}}
(\byear{2013}).
\btitle{Efficient Computer Network Anomaly Detection by Changepoint Detection
  Methods}.
\bjournal{{IEEE} Journal of Selected Topics in Signal Processing}
\bvolume{7}
\bpages{4--11}.
\bdoi{10.1109/JSTSP.2012.2233713}
\end{barticle}
\endbibitem

\bibitem{Tartakovsky+etal:SP06}
\begin{barticle}[author]
\bauthor{\bsnm{Tartakovsky},~\bfnm{Alexander~G.}\binits{A.~G.}},
  \bauthor{\bsnm{Rozovskii},~\bfnm{Boris~L.}\binits{B.~L.}},
  \bauthor{\bsnm{Bla\v{z}ek},~\bfnm{Rudolf~B.}\binits{R.~B.}} \AND
  \bauthor{\bsnm{Kim},~\bfnm{Hongjoong}\binits{H.}}
(\byear{2006}).
\btitle{A novel approach to detection of intrusions in computer networks via
  adaptive sequential and batch-sequential change-point detection methods}.
\bjournal{{IEEE} Transactions on Signal Processing}
\bvolume{54}
\bpages{3372--3382}.
\bdoi{10.1109/TSP.2006.879308}
\end{barticle}
\endbibitem

\bibitem{Tartakovsky+etal:SM06+discussion}
\begin{barticle}[author]
\bauthor{\bsnm{Tartakovsky},~\bfnm{Alexander~G.}\binits{A.~G.}},
  \bauthor{\bsnm{Rozovskii},~\bfnm{Boris~L.}\binits{B.~L.}},
  \bauthor{\bsnm{Bla\v{z}ek},~\bfnm{Rudolf~B.}\binits{R.~B.}} \AND
  \bauthor{\bsnm{Kim},~\bfnm{Hongjoong}\binits{H.}}
(\byear{2006}).
\btitle{Detection of intrusions in information systems by sequential
  changepoint methods (with discussion)}.
\bjournal{Statistical Methodology}
\bvolume{3}
\bpages{252--340}.
\bdoi{10.1016/j.stamet.2005.05.003}
\end{barticle}
\endbibitem

\bibitem{Veeravalli+Banerjee:AP2013}
\begin{bincollection}[author]
\bauthor{\bsnm{Veeravalli},~\bfnm{Venugopal~V.}\binits{V.~V.}} \AND
  \bauthor{\bsnm{Banerjee},~\bfnm{Taposh}\binits{T.}}
(\byear{2013}).
\btitle{Quickest Change Detection}.
In \bbooktitle{Academic Press Library in Signal Processing: Array and
  Statistical Signal Processing},
(\beditor{\bfnm{Rama}\binits{R.}~\bsnm{Chellappa}} \AND
  \beditor{\bfnm{Sergios}\binits{S.}~\bsnm{Theodoridis}}, eds.)
\bvolume{3}
\bpages{209--256}.
\bpublisher{Academic Press}, \baddress{Oxford, UK}.
\end{bincollection}
\endbibitem

\bibitem{Willsky+Jones:IEEE-AC1976}
\begin{barticle}[author]
\bauthor{\bsnm{Willsky},~\bfnm{Allan~S.}\binits{A.~S.}} \AND
  \bauthor{\bsnm{Jones},~\bfnm{Harold~L.}\binits{H.~L.}}
(\byear{1976}).
\btitle{A generalized likelihood ratio approach to detection and estimation of
  jumps in linear systems}.
\bjournal{{IEEE} Transactions on Automatic Control}
\bvolume{21}
\bpages{108--112}.
\bdoi{10.1109/TAC.1976.1101146}
\end{barticle}
\endbibitem

\bibitem{Woodroofe:Book82}
\begin{bbook}[author]
\bauthor{\bsnm{Woodroofe},~\bfnm{Michael}\binits{M.}}
(\byear{1982}).
\btitle{Nonlinear Renewal Theory in Sequential Analysis}.
\bpublisher{Society for Industrial and Applied Mathematics},
  \baddress{Philadelphia, PA}.
\end{bbook}
\endbibitem

\bibitem{Zhigljavsky:IWSM2009}
\begin{binproceedings}[author]
\bauthor{\bsnm{Zhigljavsky},~\bfnm{Anatoly}\binits{A.}}
(\byear{2009}).
\btitle{Application of the {S}ingular {S}pectrum {A}nalysis for Change-point
  Detection in Time Series}.
In \bbooktitle{Proceedings of the 2nd {I}nternational {W}orkshop in
  {S}equential {M}ethodologies}.
\end{binproceedings}
\endbibitem

\bibitem{Zhigljavsky+Kraskovsky:Book1988}
\begin{bbook}[author]
\bauthor{\bsnm{Zhigljavsky},~\bfnm{Anatoly~A.}\binits{A.~A.}} \AND
  \bauthor{\bsnm{Kraskovsky},~\bfnm{Alexander~E.}\binits{A.~E.}}
(\byear{1988}).
\btitle{Detection of Abrupt Changes of Random Processes in Radiotechnics
  Problems}.
\bpublisher{St. Petersburg University Press}, \baddress{St. Petersburg,
  Russia}.
\bnote{(in Russian)}.
\end{bbook}
\endbibitem

\end{thebibliography}

\end{document}